\documentclass[pdflatex,sn-mathphys-num]{sn-jnl}

\usepackage{graphicx}%
\usepackage{multirow}%
\usepackage{amsmath,amssymb,amsfonts}%
\usepackage{amsthm}%
\usepackage{mathrsfs}%
\usepackage[title]{appendix}%
\usepackage{xcolor}%
\usepackage{textcomp}%
\usepackage{hyperref}
\usepackage{manyfoot}%
\usepackage{booktabs}%
\usepackage{algorithm}%
\usepackage{algorithmicx}%
\usepackage{algpseudocode}%
\usepackage{listings}%
\usepackage{csquotes} %
\usepackage{xr} %
\usepackage{bm} %
\usepackage{multibib}
\newcites{SI}{SI References}

\theoremstyle{thmstyleone}%
\theoremstyle{thmstyletwo}%

\theoremstyle{thmstylethree}%

\begin{document}
\title[Article Title]{A Unified Predictive and Generative Solution for Liquid Electrolyte Formulation}


\author[1]{\fnm{Zhenze} \sur{Yang}}
\author[1,**]{\fnm{Yifan} \sur{Wu}}
\author[1]{\fnm{Xu} \sur{Han}}
\author[1]{\fnm{Ziqing} \sur{Zhang}}
\author[1,**]{\fnm{Haoen} \sur{Lai}}
\author[1]{\fnm{Zhenliang} \sur{Mu}}
\author[1]{\fnm{Tianze} \sur{Zheng}}
\author[1]{\fnm{Siyuan} \sur{Liu}}
\author[1]{\fnm{Zhichen} \sur{Pu}}
\author[1]{\fnm{Zhi} \sur{Wang}}
\author[1]{\fnm{Zhiao} \sur{Yu}}
\author[1,*]{\fnm{Sheng} \sur{Gong}}
\author[1,*]{\fnm{Wen} \sur{Yan}}


\affil[1]{\orgdiv{ByteDance Seed}}

\affil[**]{work done as intern at ByteDance Seed}

\affil[*]{corresponding: sheng.gong@bytedance.com, wen.yan@bytedance.com}

\abstract{Liquid electrolytes are critical components of next-generation energy storage systems, enabling fast ion transport, minimizing interfacial resistance, and ensuring electrochemical stability for long-term battery performance. However, measuring electrolyte properties and designing formulations remain experimentally and computationally expensive. In this work, we present a unified framework for designing liquid electrolyte formulation, integrating a forward predictive model with an inverse generative approach. Leveraging both computational and experimental data collected from literature and extensive molecular simulations, we train a predictive model capable of accurately estimating electrolyte properties from ionic conductivity to solvation structure. Our physics-informed architecture preserves permutation invariance and incorporates empirical dependencies on temperature and salt concentration, making it broadly applicable to property prediction tasks across molecular mixtures. Furthermore, we introduce---to the best of our knowledge---the first generative machine learning framework for molecular mixture design, demonstrated on electrolyte systems. This framework supports multi-condition–constrained generation, addressing the inherently multi-objective nature of materials design. As a proof of concept, we experimentally identified three liquid electrolytes with both high ionic conductivity and anion-concentrated solvation structure. This unified framework advances data-driven electrolyte design and can be readily extended to other complex chemical systems beyond electrolytes.}




\maketitle

\section*{Introduction}\label{sec1}

Liquid electrolyte is one of the most important components in both the current commercial lithium ion batteries~\cite{xu2004nonaqueous,meng2022designing} and the next-generation lithium-metal batteries~\cite{wang2022liquid}. 
Different solvent molecules and salts have been proposed as components of liquid electrolytes with corresponding advantages, such as cyclic carbonates with strong solvating power to Li$^{+}$, linear carbonates, esters and ethers with relatively low viscosity and moderate solvating power~\cite{xu2004nonaqueous,meng2022designing}, and various fluorinated molecules for LiF-dominated solid-electrolyte-interphase (SEI)~\cite{li2023critical,yu2020molecular}. 
In this context, the formulation of electrolytes plays a critical role in optimizing battery performance~\cite{chen2025hybrid}, yet developing novel formulations remains challenging due to the vast chemical space of electrolyte molecules and the intractable number of possible molecular combinations. 
Despite extensive research, experimental measurements for key properties such as ionic conductivity and coulombic efficiency remain sparse, with approximately 10,000 conductivity data points available across various public datasets~\cite{Kumar2024electrolytomics, Blasio2024, Bradford2023} and only a few hundred for coulombic efficiency~\cite{Kim2023}. 
These limited data cover only a small fraction of the compositional space of potential electrolytes, significantly hindering the efficient exploration of novel formulations. 

To supplement experimental measurements, MD simulations using classic force fields such as OPLS series~\cite{Jorgensen1996,Doherty2017} have been widely employed to estimate liquid electrolyte properties. 
Despite the low computational cost, these classical force fields often struggle to achieve satisfactory prediction accuracy for properties of liquid electrolytes~\cite{Gong2025}, due to the simplicity of the functional forms of classic force fields.
While machine learning force fields (MLFFs)~\cite{Wang2023,Dajnowicz2022,Gong2025} might offer improved accuracy over classic force fields, the computational cost of MLFF remains significantly higher, making MLFF-based large-scale screening impractical. 

Recent advancements in data-driven machine learning (ML) approaches have shown promise in directly predicting electrolyte properties from molecular mixtures. 
Several studies have successfully employed deep learning (DL) architectures, such as graph neural networks (GNNs) and attention-based models, to achieve reasonable accuracy, demonstrating that ML models can extract meaningful patterns from limited experimental datasets. 
For instance, ~\citet{Zhang2024Molset} proposed the \enquote{MolSet} model for predicting the conductivity of lithium battery electrolytes, while ~\citet{Zeng2024} leveraged multi-level information to capture various electrolyte formulation properties. 
However, these models still face two major limitations: limited formulation coverage and lack of physical constraints. 
For instance, the \enquote{MolSet} work is limited to handling mixtures containing up to four different molecule types and primarily focuses on polymer electrolytes, and that from~\citet{Zeng2024} does not include physical constraints, which might lead to occurrences of unphysical results for property prediction. Besides experimental data, \citet{Chew2025} also utilized high-throughput molecular simulations together with ML models for property predictions of chemical mixtures. However, a significant gap remains between the results of molecular simulations and experimentally measured bulk properties.

In addition to the latest progress in ML-based property prediction models for liquid electrolytes—often regarded as a \enquote{forward problem}—an equally practical and impactful challenge lies in the \enquote{inverse problem}: how to efficiently explore and design novel electrolyte formulations that satisfy specific property requirements. 
This task becomes particularly challenging given the vast design space of electrolyte formulations. 
For instance, given a formulation where eight components are selected from a pool of 100 candidate molecules, and the molar ratios of each molecule range from 0 to 1 with an interval of 0.05, the number of possible combinations reaches approximately $C_{100}^{8} \times C_{27}^{7} \approx 1.65 \times 10^{17}$ (combinations of molar ratios can be reformulated as a non-integer composition problem). 
The diversity of molecular species, combined with the variability in molar ratios, results in an expansive design space for electrolyte formulations. 
In the current commercial electrolytes, electrolyte formulations with more than five, and in some cases, up to ten different component molecules~\cite{Faustov2021,Cheng2016} are commonly used to meet the diverse performance requirements of a commercial battery cell. 
Electrolytes composed of molecular mixtures often exhibit enhanced properties, such as higher conductivity and coulombic efficiency, compared to single-molecule systems~\cite{chen2025hybrid}. 
Moreover, the concept of high-entropy electrolytes, which are composed of a diverse set of solvents and salts, is gaining increasing attention for their potentially high ionic conductivity, cycling stability, and rate capacity~\cite{Wang2023natcomm, Kim2023natenergy, Wang2024}. 
These examples highlight the vast design space of electrolyte formulations, emphasizing the need for efficient exploration methods. However, most existing data-driven studies for liquid electrolyte focus on local optimization rather than global exploration. 
For instance, ~\citet{Zhu2024} optimized only ternary mixtures, while ~\citet{Zeng2024} primarily screened single solvent molecules to boost conductivity of electrolytes. 
As a result, these models, while effective in solving their respective designated tasks, do not naturally scale to the full complexity of multi-component electrolyte formulation.

To address these challenges, generative models offer a compelling alternative for systematically navigating the electrolyte design space. 
Unlike brute-force screening methods or local optimization approaches, generative approaches can propose novel electrolyte formulations or molecular mixtures by directly learning from the data distribution, making them well-suited for high-dimensional search spaces~\cite{Taylor2022, Liu2023}. 
\textbf{However, to the best of our knowledge, no prior work has applied generative modeling to molecular mixture generation, letting alone specific property-guided electrolyte generation.} 
This is in stark contrast to other material design fields, such as molecular discovery~\cite{anchez-Lengeling2018, Du2024, Bilodeau2022} and crystal structure generation~\cite{Xie2022, Zeni2025,Anstine2023}, where generative models have been widely adopted. 
This gap exists because of several fundamental challenges:
\begin{itemize}
    \item \textbf{Large and Structured Design Space} --- Molecular mixture design involves a vast chemical space due to the diversity of molecular species and the wide range of possible stoichiometric mixing ratios. This combinatorial complexity significantly enlarges the design space compared to single-component systems.
    \item \textbf{Complex Interactions} --- Even given a fixed set of chemicals and a specific mixing ratio, accurately simulating mixture properties remains highly challenging due to intricate intermolecular and intramolecular interactions. As a result, data-driven predictions often rely heavily on experimental measurements.
    \item \textbf{Data Scarcity} --- Despite their importance, open-sourced experimental data for molecular mixtures are extremely limited due to the high cost and time required for synthesis and characterization, as well as the commercial interest of molecular mixtures. Computational data is also expensive to obtain at scale.
    \item \textbf{Representation and Inductive Bias} --- Effective methods like ML models for molecular mixtures need to incorporate appropriate inductive biases and representations that reflect the underlying physics, such as permutation invariance with respect to molecular ordering, to address the challenges of scale, complexity, and data scarcity.
\end{itemize} 
These challenges highlight the need for a novel approach that integrates physics-aware representations~\cite{Karniadakis2021}, data-driven conditional generative modeling, and efficient evaluation strategies. 
In this work, we develop a unified framework for electrolyte formulation design, integrating a forward predictive model and an inverse generative approach. 
The main contributions of this paper are summarized as follows:
\begin{itemize}
    \item We extensively collect literature data for both single molecules (240,000+) and molecular mixtures (10,000+) with labeled properties, enabling broad coverage of the electrolyte design space. 
    By further integrating over 100,000 molecular mixture data points generated from molecular dynamics (MD) simulations, we are able to train an accurate ML model for not only conductivity prediction, but also solvation structure estimation which might relate to interfacial stability of Li metal batteries.
    \item We propose a physics-informed architecture that preserves permutation invariance and incorporates empirical dependencies on temperature and salt concentration for accurate property prediction. 
    This approach serves as a universal framework applicable to a wide range of prediction and design tasks involving molecular mixtures, extending beyond electrolyte formulations.
    \item We develop the first generative framework for molecular mixture design, using electrolyte systems as a representative example. 
    We introduce a multi-condition–constrained generative approach, providing a promising solution for multi-objective materials design.
    \item We further tested generated electrolyte formulations with experimental conductivity measurement and Raman spectroscopy. Among these promising candidates, we identified three formulations that exhibit both high ionic conductivity and anion-concentrated solvation structure.
\end{itemize}

\section*{Results}\label{sec:results}
\subsection*{Overall workflow}
The overall workflow of this study is illustrated in Fig.~\ref{fig:overview}a and comprises two key components: a forward predictive model and an inverse generative process. 
The predictive model estimates two target properties—ionic conductivity and coordination ratio of anion around Li$^+$ (abbreviated as anion ratio below)—based on various electrolyte formulations. 
In contrast, the generative process designs new electrolyte formulations conditioned on desired property values. 
Both components follow three-stage training and evaluation procedures, described in detail below.  

In the forward prediction process, we first employ a GNN model to learn a universal molecular embedding (also referred to as a \enquote{fingerprint} or \enquote{descriptor}) for each molecule within the electrolyte system. 
This embedding is learned through a multi-task pretraining strategy, using a large single-molecule dataset (over 240,000 entries) to predict various molecular properties such as melting point (Process (1) in Fig.~\ref{fig:overview}b). 
Next, the molecular embeddings of electrolyte components are aggregated—weighted in a learnable way by their molar ratios—into a single electrolyte-level embedding using a permutation-invariant neural network. 
To broaden coverage of the formulation space, this \enquote{Invariant Aggregation} model is pretrained on over 100,000 electrolyte systems calculated via MD simulations (Process (2) in Fig.~\ref{fig:overview}b). 
Finally, the model is fine-tuned on more than 10,000 experimental electrolyte data points (62 solvents and 17 Li salts) and incorporates empirical relations with physical priors to enhance conductivity prediction accuracy (Process (3) in Fig.~\ref{fig:overview}b).

The generation process can be viewed as the inverse of the prediction task. We begin by training a conditional diffusion model to generate electrolyte embeddings given specified property targets—namely, anion ratio and conductivity (Process (1) in Fig.~\ref{fig:overview}c). 
These generated electrolyte embeddings are then converted back to a set of molecular embeddings and corresponding molar ratios using a trainable decoder model (Process (2) in Fig.~\ref{fig:overview}c). 
Finally, the generated molecular embeddings are matched to specific molecules in our electrolyte database based on the distance of embeddings (Process (3) in Fig.~\ref{fig:overview}c). 
The selected molecules and their corresponding molar ratios together define the final generated electrolyte formulation. 

The detailed model architectures within the whole workflow are visualized in Fig.~\ref{si_fig:model_architecture} and data representations of molecules and electrolytes are summarized in Table~\ref{si_tab:data_repr}. More information will be discussed in the following sections and Supplementary Information.
\subsection*{Electrolyte property prediction}
In this section, we discuss in detail about the methodologies and results of the forward predictive model in our workflow. 
\subsubsection*{Molecular pretraining}
To obtain a comprehensive and efficient representation of diverse molecules within the electrolyte formulations, we pretrained a GNN model to learn a universal \enquote{fingerprint} by predicting various molecular properties with more than 200,000 molecular data. 
These properties include 11 distinct entries: melting point ($T_m$), boiling point ($T_b$), (liquid) refractive index ($n_D$/$n_D^{\text{liquid}}$), p$K_a$, p$K_b$, dielectric constant ($\varepsilon$), surface tension ($\gamma_s$), density ($\rho$), viscosity ($\eta$) and vapor pressure ($P_{vap}$). 
The GNN model takes atomic and bond features derived from the chemical SMILES representation~\cite{Weininger1988} as input and outputs a molecular embedding using a modified Edge-augmented Graph Transformer (EGT) model~\cite{Zheng2025,Hussain2022}. 
This embedding is then processed through separate readout blocks with multiple multilayer perceptrons (MLPs), each dedicated to one respective molecular properties.  

The results of molecular pretraining are shown in Fig.~\ref{fig:prediction}a. 
We compare our GNN architecture (labeled as \enquote{Atom/bond feature + EGT}) with two baseline methods: \enquote{Morgan fingerprint + NN} and \enquote{Atom/bond feature + GAT} (see the \hyperref[sec:molecular_pretraining]{Methods} section for details). 
Our GNN model consistently outperforms both baselines in the prediction of all molecular properties and achieves high $R^2$ scores across these tasks (see also in Fig.~\ref{si_fig:pretrain_scatter}). 
These findings suggest that the molecular embeddings learned during pretraining encode rich chemical information, enabling their effective application in downstream tasks on electrolytes, which are mixtures of the constituent molecules. 

\subsubsection*{Physics-informed architecture}
With a comprehensive descriptor for each individual molecule, the next step is to understand how they collectively influence the properties of liquid electrolyte systems. 
To this end, we developed an invariant aggregation block, followed by an empirical equation block, to integrate molecular information and predict electrolyte properties with physical prior. 

More specifically, our model incorporates two key physical constraints: permutation invariance and the dependence of conductivity on temperature and concentration.  
Unlike predictive or generative tasks involving a single molecule, an electrolyte formulation is a mixture of multiple molecules, and its properties should remain invariant under permutations of its components. 
To achieve this, we employ a self-attention-based model that performs a learnable, weighted aggregation of molecular embeddings, guided by the molar ratios of the constituent molecules in the formulation (see \hyperref[sec:perm_inv]{Methods} and Supporting Information for more details). 
The aggregation block produces an electrolyte-level embedding, which is then used to predict the parameters of our empirical equation that models the dependence of conductivity on temperature and concentration. 

In terms of empirical relation, dependence of electrolyte conductivity on temperature and salt concentration has been extensively studied, with a large number of theoretical or empirical models~\cite{Kontogeorgis2018, Casteel1972, ngai2011relaxation, Fu2018, Zhang2020}. 
Some of these general trends are well-known. 
For instance, conductivity typically increases initially and then decreases with increasing salt concentration due to the trade-off between the number of charge carriers and the increasing viscosity. 
In addition, conductivity increases with rising temperature due to enhanced ion mobility. 
To incorporate these domain knowledge, prevent unphysical prediction (Fig.~\ref{si_fig:outlier}) and enhance generalizability, we employed an empirical equation (Eq.~\ref{eq:empirical_equation}) for the temperature and concentration dependence of conductivity based on previous studies~\cite{Fu2018, Zhang2020} and our evaluation (Fig.~\ref{si_fig:emp_relation}, see \hyperref[sec:empirical_relation]{Methods} section and Supporting Information for the derivation, mathematical form and ablation study of the empirical equation). 

With the empirical equation, model inference is no longer required each time the temperature or concentration change. 
Instead, the entire temperature and concentration curve for an electrolyte system can be obtained with one-time inference of our model, which dramatically improves the efficiency of conductivity prediction. 
The peak conductivity and its corresponding salt concentration can also be directly computed using the empirical relation. 
In addition, we also incorporated viscosity in the empirical relation as conductivity generally increases with decreasing viscosity. 
We utilized an inverse relation between conductivity and viscosity described by the well-known Walden's rule~\cite{Zhang2022}. 
The incorporation of viscosity into the empirical relation further enhances the generalizability of our model using the viscosity as a prior for conductivity prediction (see Table~\ref{si_tab:generalization}, data split based on viscosity). 

\subsubsection*{Pretraining with computational data}
To enable broader coverage of the electrolyte design space and capture solvation characteristics, we pretrained our electrolyte-level model using data generated from MD simulations (denoted as \enquote{computational pretraining}). 
Specifically, we conducted over 100,000 all-atom MD simulations using the OPLS force field~\cite{Jorgensen1996,Doherty2017} to obtain key electrolyte properties, including ionic conductivity and anion ratio. 
High ionic conductivity enables rapid ion transport, which minimizes internal resistance, reduces energy loss, and supports high power output and fast charging. 
However, a well-known trade-off exists between ionic conductivity and interfacial stability of liquid electrolytes used in lithium-based batteries. 
This trade-off arises because organic solvents that provide high ionic conductivity---such as carbonates or ethers---tend to have low electrochemical and chemical stability, especially when in contact with reactive electrode materials like lithium metal or high-voltage cathodes~\cite{Xu2004,Tikekar2016}. 
Recent works addressed this challenge in the lithium-metal battery by reducing the free solvent molecules within the Li$^+$ solvation structure (known as \enquote{weak solvation}), leading to a predominantly inorganic SEI for better Li cyclability~\cite{Yu2020, Li2023, Chen2024, Wu2024c, Emilsson2025}. 
Inspired by these studies, we use the anion ratio---which correlates positively with Li cyclability---as an additional objective alongside conductivity. Here, we define the anion ratio as the number of anions divided by the total number of anions and solvent molecules in the Li ion’s first solvation shell. A larger anion ratio results in weaker solvation of solvents, thereby generally enhancing the battery’s interfacial stability. 

Given the extensive MD data, computational pretraining not only allows us to obtain relevant information to Li cyclability, but also provides broader coverage of the electrolyte formulation design space—particularly for multi-component systems. 
Experimental studies in the literature typically focus on liquid electrolytes with only a few solvents and a single type of Li salt. 
In addition, incorporating a new molecule into experiments is generally more costly---and sometimes impractical—compared to MD simulations. 
Consequently, computational pretraining serves as a valuable tool to improve predictive accuracy when exploring novel chemical candidates for electrolyte design as we show in Table~\ref{si_tab:generalization}. 

To exploit diverse information from MD simulations, pretraining is conducted in a multi-task learning manner, simultaneously predicting anion ratio and ionic conductivity. 
The conductivities are obtained from two different methods: Mistry’s method~\cite{Mistry2023} (denoted as \enquote{mistry}) and Nernst-Einstein (denoted as \enquote{NE}) method (see \hyperref[sec:comp_pretrain]{Methods} section and Supplementary Information for more details). 
Although there are some simplifications in these two methods (NE ignores the correlation between ions and Mistry's method calculates chemical potential under dilute assumption), we observe positive correlation between MD-calculated and experimental conductivities and consequently believe that MD data remains useful for the pretraining task (Fig.~\ref{si_fig:cond_compare}). 
The prediction results of these three properties are plotted in Fig.~\ref{si_fig:comp_pretrain}. 
We visualized the correlation between anion ratio and conductivity derived from MD data in Fig.~\ref{fig:prediction}b, which displays a clear trade-off between these two properties. 

\subsubsection*{Fine-tuning with experimental data}
Following computational pretraining on extensive MD simulation data, we further fine-tune our model using over 10,000 experimentally measured conductivity values collected from the scientific literature (denoted as \enquote{experimental fine-tuning}). 
During fine-tuning, the molecular embeddings learned from pretraining are kept fixed, while the electrolyte embeddings are updated. 
Since experimental data for anion ratio are not available, we continue to use MD-derived anion ratios during this stage. For conductivity prediction, our model estimates the electrolyte embedding‑dependent parameters of the empirical relation. Conversely, because no empirical relation exists for anion ratio, we concatenate the electrolyte embedding with temperature and concentration and employ a readout layer to predict the property.
As shown in Fig.~\ref{fig:prediction}c, the fine-tuned model accurately predicts both experimental ionic conductivity and computational anion ratio, achieving $R^2$ = 0.985 for conductivity and $R^2$ = 0.953 for anion ratio. 

We further examine the temperature and concentration dependence of conductivity predicted by the model. 
As Fig.~\ref{fig:prediction}d reveals, the model precisely reproduces both trends across a wide range of electrolyte systems. 
The temperature dependence typically follows a linear relationship between log-scale conductivity and $\frac{1}{T - T_0}$, where $T_0$ is a system-dependent parameter generally associated with the glass transition temperature of the electrolyte~\cite{ngai2011relaxation}. 
In contrast, the concentration dependence exhibits a characteristic \enquote{volcano} plot, with conductivity first increasing and then decreasing as concentration changes. 
These trends are rigorously enforced through the incorporation of our empirical relation. More examples of temperature and concentration dependence of conductivity can be found in Fig.~\ref{si_fig:tmdep}. These results demonstrate that our adapted empirical relations are applicable across different electrolyte systems and that our model accurately captures the effects of experimental conditions. 

\subsection*{Generative electrolyte design}
In this section, we discuss in detail about the methodologies and results of the inverse generative model in our workflow.

\subsubsection*{Conditional generation}
The forward predictive model enables rapid evaluation of electrolyte properties, making it an efficient tool for screening electrolyte formulations. 
However, as the candidate space grows exponentially as the number of molecule species in the mixture formulation, brute-force sequential screening becomes intractable to design electrolyte systems with target properties. 
As an alternative, we leverage a diffusion model, to conditionally generate electrolyte formulations that satisfy specific property targets. 

We utilize a synthetic dataset generated by our predictive model to train our conditional diffusion model. 
The conditional diffusion model takes conductivity and anion ratio as constraints and generates electrolyte embeddings from random Gaussian noises (see \hyperref[sec:cond_diff]{Methods}). 
The generated electrolyte embeddings are converted back to molecular embeddings using a decoder module (Fig.~\ref{si_fig:model_architecture}e). 
Each molecular embedding is then matched to molecules in our electrolyte database based on a minimum distance criterion (see Supplementary Information for more details). 
Given a fixed set of molecular species, an electrolyte formulation can be represented as a normalized vector, referred to as the \enquote{Bag of Molecules} (BoM) vector. 
Each dimension of the \enquote{BoM} vector corresponds to the molar ratio of the molecule at that position. 
The concept is similar as the \enquote{Bag of Words} representation in Natural Language Processing (NLP). 
With the formulation representation, we further utilized our predictive model to evaluate the properties of generated electrolyte formulations. 

The results of conditional generation for electrolyte formulations are plotted in Fig.~\ref{fig:generation}. 
As shown by Fig.~\ref{fig:generation}a, the generative model enables the design of electrolyte formulations with extrapolated properties in both conductivity and anion ratio. 
By conditioning the diffusion model on high conductivity or anion ratio values, the resulting data distribution is significantly shifted toward the desired property targets. 
In addition, we highlight that the model can generate electrolyte formulations that simultaneously satisfy both property targets. For example, under the condition of 10.0 mS/cm conductivity and an anion ratio of 0.3, the model’s output is shown in Fig.~\ref{fig:generation}b, and a representative formulation appears in Fig.~\ref{fig:generation}c. We further evaluate conditional generation across conductivity constraints ranging from 5 to 30 mS/cm and anion ratios from 0.1 to 0.7; these results are visualized in Fig.~\ref{si_fig:cond_generation}, demonstrating the model’s versatility.
More generated samples under various conditions can be found in Fig.~\ref{si_fig:example_formulation_5.0} through Fig.~\ref{si_fig:example_formulation_30.0}. 
These results demonstrate the strong capability of the generative model, which can largely enhance the efficiency of designing electrolytes with target properties.

\subsubsection*{Performance evaluation}
To further quantify the performance of generation, we proposed three different evaluation metrics including one accuracy score (mean absolute percentage error, MAPE hereafter) and two diversity scores. 
The MAPE score is calculated by summarizing the relative error from both conductivity and anion ratio. 
In contrast, the two diversity metrics assess the similarity among generated formulations to help prevent mode collapse, a common issue in generative modeling. 
These two diversity scores highlight different aspects: formulation diversity measures the average pairwise distance between closest formulations within the generated set, while molecular diversity evaluates the entropy of molecular occurrence across generated samples (Mathematical definitions of diversity scores are discussed in Supporting Information). 

With these evaluation metrics, we show that in Fig.~\ref{fig:generation}d, both accuracy and diversity are higher when the generation conditions are close to the original labeled data distribution. 
In our synthetic dataset, the mean conductivity is approximately 5.57 mS/cm and the mean anion ratio is 0.269. 
Accordingly, generation conditioned on conductivity = 5.0 mS/cm and anion ratio = 0.3 yields low error and high diversity. 
Conversely, when the target properties lie at the tails of the data distribution (e.g. conductivity = 30.0 mS/cm and anion ratio = 0.7), the model tends to produce electrolyte formulations with larger deviations from the target properties and reduced diversity. 
In addition, there is a general trade-off between the generation accuracy and diversity shown by Fig.~\ref{fig:generation}d. 

\subsubsection*{Electrolyte generation under base formulation constraints}
In addition to the two target properties of interest, incorporating a base formulation as a constraint is also one of the common practices in electrolyte formulation design. 
A base formulation poses constraints on the molar ratios of certain molecules. 
The motivations for this are as follows:
\begin{itemize}
    \item Incorporating base formulation constraints helps reduce the design space of electrolyte formulations, thereby improving the efficiency of the design process.
    \item Practical battery design often requires electrolytes to meet a variety of additional constraints. 
    By using a base formulation, we can integrate prior design knowledge beyond the specific conditions imposed on the diffusion model. 
    For example, the molar ratio of ethylene carbonate (EC) in the solvent mixture is typically maintained above 20\% to enhance the solubility of Li salts, while fluoroethylene carbonate (FEC) is usually kept below 10\% in industrial applications due to cost considerations.
    \item This approach becomes especially important when certain property requirements—such as electrochemical stability window, interfacial compatibility, or thermal stability—are expensive and labor-intensive to measure. 
    Consequently, the available data for training ML models can be limited. 
    Applying base formulation constraints can help mitigate this issue by guiding the generation process within a more informed and feasible region of the design space.
\end{itemize}
Therefore, we here realize the conditional generation with the base formulation constraints by utilizing a classifier-guided diffusion (CGD) approach~\cite{Cgd2021}(Details can be found in \hyperref[sec:cond_diff]{Methods} and Supporting Information).  
Specifically, during the sampling process, we use our decoder module as a classifier to convert noisy electrolyte embeddings into \enquote{BoM} vectors, which are then evaluated to determine whether the generated formulations satisfy the base formulation constraints. 
The gradient from the classifier is subsequently used to guide the denoising process toward the desired base formulation constraints. 
A schematic of this classifier-guided sampling is displayed in Fig.~\ref{fig:generation}e. 
This approach offers several advantages over conventional classifier-guided or classifier-free diffusion methods. 
Above all, the decoder is computationally inexpensive, so incorporating gradient calculations during sampling does not really increase the overall sampling time. 
Additionally, there is no need to retrain the diffusion model or classifier each time a new constraint is introduced. 
The decoder module only needs to be trained once, and adapting to different base formulation constraints simply requires redefining the classification criteria on the decoded formulations. 
This significantly improves the efficiency and flexibility of the generation process. 

We tested our CGD approach on two different base formulation constraints: (1) EC $>$ 20\%; (2) EC $>$ 20\% \& DMC (dimethyl carbonate) $>$ 20\% \& EMC (ethyl methyl carbonate) $>$ 20\%. 
For conditional generation, the target conductivity was set to 10 mS/cm and the anion ratio to 0.25, and more than 10,000 samples are generated for evaluation. 
As shown in Fig.~\ref{fig:generation}f, our CGD method significantly improves the success rate of electrolyte formulation generation compared to the case without classifier guidance. 
For case (1), the success rate is improved by a factor of 30, while for case (2), it is enhanced by at least three orders of magnitude (without CGD method, there is no single formulation meets the condition among more than 10,000 generated samples). 
The success rate accounts not only for satisfying the base formulation constraints, but also for generating electrolytes with conductivity and anion ratio values close to the specified targets. 
These results demonstrate that CGD can effectively guide the generative process toward realistic, multi-constraint electrolyte formulations with high fidelity.

\subsection*{Experimental validation}
As a proof of concept, we further conducted experiments to validate the properties of the generated formulations. 
Specifically, we measured ionic conductivity and used Raman spectroscopy to probe the solvation structure of electrolytes. 
Although experimentally quantifying ratio of anion coordination is not straightforward, Raman spectroscopy can provide insights into the solvation structure by probing the degree of coordination between anion and cation in the solution~\cite{lai2025linking}.

To select formulations for experimental validation, we first applied our conditional generative workflow to propose candidates targeting conductivities of 15 or 20 mS/cm together with anion ratios of 0.3 or 0.5. 
These generated formulations were then screened using our predictive models to refine the selection. 
Those that have minimal deviation from the target properties were selected. Given the availability of molecules from commercial chemical vendors, expert inspection for stability and solubility, experimental costs and data validity, we selected 18 formulations for experimental validation including 7 reference formulations with commonly seen single solvent (PC, EC, THF, EA, MA, DEC, DMC) and 11 generated formulations with multi-solvents. 

Among the 11 generated formulations, 10 mixture formulations were experimentally validated with ionic conductivity within the target range, significantly outperforming the average conductivity of randomly selected experimental formulations (Fig.~\ref{fig:exp_validation}a). 
More importantly, Raman spectroscopy analysis reveals a clear rightward shift of the characteristic FSI$^-$ peak (typically observed between 710 and 730 cm$^{-1}$) in three high-conductivity generated formulations (around 15 mS/cm) compared to typical single-solvent electrolytes (Fig.~\ref{fig:exp_validation}b). Among these single-solvent electrolytes, DEC and DMC are well-studied for promoting anion-concentrated solvation structures~\cite{Fan2024, lai2025linking}.
These spectral shifts suggest a high degree of anion-cation coordination, indicating that in these generated high-conductivity electrolytes FSI$^-$ anions are more likely to enter the first solvation shell of Li$^+$. 
This behavior corresponds to a higher anion ratio and might lead to enhanced interfacial stability. 
Notably, the formulation \enquote{MA/DEC/EA/EMC/LiFSI} exhibits a right shift of the FSI$^-$ Raman peak compared to each individual solvent component (EMC and DMC have similar FSI$^-$ peak positions~\cite{su2019solvating}), suggesting a solvation structure beyond interpolation of each individual component within the electrolyte. 
More details about the compositions of solvents and salts within each formulation can be found in Table~\ref{si_tab:mol_name} and Table~\ref{si_tab:exp_formulations}. 
These experimental results demonstrate that our generative workflow can propose novel and intriguing electrolyte formulations that exhibit both high conductivity and potentially high interfacial stability.

\section*{Discussion}\label{sec:discussion}
In summary, this study presents a unified framework that integrates both predictive and generative tools for electrolyte formulation. 
By leveraging a combination of computational and experimental data—spanning single molecules to molecular mixtures—the predictive model enables accurate property estimation from the atomic scale to the formulation level. 
Incorporating physics-informed priors, the model's prediction not only preserves permutation invariance, but also observes the typical temperature and concentration dependencies of ionic conductivity observed in electrolyte systems. 
Furthermore, through generative modeling, we significantly enhance the efficiency of electrolyte design and enable multi-objective generation, allowing simultaneous control over key formulation properties and constraints. 
This capability is particularly valuable, as practical electrolytes often consist of complex mixtures of multiple component molecules and must meet a range of performance criteria. 
Our approach provides a scalable and flexible framework to navigate this multi-dimensional design space more effectively.

However, there remains substantial future work to address the current limitations of the workflow.
For instance, given the limited data availability—such as only a few hundred data points for Coulombic efficiency-we use anion ratio as an indirect proxy for evaluating the interfacial stability of electrolytes.
This property is not as direct an indicator of battery cyclability as Coulombic efficiency.
However, as data of battery performance becomes increasingly accessible, this framework can be readily extended to incorporate additional performance metrics, such as electrochemical stability window, Li-ion transference number, and thermal stability~\cite{Wang2022, Teng2023}, offering more direct insights into electrolyte behaviors within battery systems.

In addition, while the current workflow operates within a fixed molecular space, our model has the capacity to extend to a much broader chemical space.
This is made possible by the decoder architecture, which recovers molecular embeddings before molecule matching.
These embeddings can either be matched against a larger chemical database to identify promising candidates, or used as input to a molecular decoder trained to reconstruct molecular structures in the form of SMILES strings or 2D molecular graphs for molecular discovery~\cite{Gmez-Bombarelli2018, Bagal2022}. 

Last but not least, the methodology developed in this work for property prediction and conditional generation of liquid electrolytes with target properties is potentially transferable to other systems composed of mixtures of chemical species.
These systems include but not limited to fossil fuels~\cite{Nursulu2022}, biomolecular aggregates~\cite{Abramson2024}, high-entropy alloys~\cite{Easo2019}, polymer blends~\cite{Ge2025}, ionic liquids~\cite{Koutsoukos2021}, deep eutectic solvents~\cite{Hansen2021}, and multi-component catalysts~\cite{SchlexerLamoureux2019}.
Given that current research efforts in the overall AI4Science field largely focus on the prediction and generation of mono-material systems—such as protein monomers, single crystals, or individual molecules—we believe there is substantial opportunity to adapt our workflow to a wide range of complex, multi-component material systems in the future.

\section*{Methods}\label{sec:methods}

\subsection*{Molecular pretraining}\label{sec:molecular_pretraining}
The single-molecule dataset used for molecular pretraining was compiled from various public sources (Table~\ref{si_tab:mono_dataset}) and carefully screened to eliminate ambiguous or incorrect data. 
The final dataset comprises a total of 189,528 unique molecules with 241,414 entries under different temperatures.
Among 11 molecular properties we collect, density, refractive index, dielectric constant, surface tension, viscosity and vapor pressure are temperature-dependent (More details can be found in the Supplementary Information). For temperature-dependent properties, each unique temperature-property pair is treated as an independent data point for training. \\
\\
For predicting molecular properties, we implement a GNN architecture, closely following the graph block design of the ByteFF model~\cite{Zheng2025}. 
This architecture leverages both atom features (element type, ring connectivity, minimum ring size, and formal charge) and bond features (bond order and ring membership) as molecular input representations. 
To effectively create a molecular descriptor, a modified EGT is incorporated within the GNN, enabling the model to learn node and edge embeddings from atomic and bonding information. 
The final molecular embedding is constructed by concatenating the node and edge embeddings generated by the EGT block. 
Further in multi-property prediction, the universal molecular embedding is processed through separate readout blocks, establishing a multi-task learning framework that enables simultaneous prediction of multiple molecular properties. 
For those properties that are temperature-dependent, we adopt different empirical relations and use readout layers to predict parameters in these equations to obtain corresponding properties (see Supporting Information for more details). 
To benchmark the performance of this GNN-based approach and obtain an effective molecular embedding, we compare the performance with two baseline models including: (1) a simple feed forward neural network using Morgan fingerprint~\cite{Morgan1965} as input (\enquote{Morgan fingerprint + NN}) or (2) a Graph Attention Network (GAT)~\cite{velickovic2018graph} as an alternative to EGT using node/atom features as input (\enquote{Atom/bond feature + GAT}). 
As shown in Table~\ref{si_tab:pretrain_comparison}, the selected EGT model outperforms the two baseline models, achieving the highest accuracy across nearly all property prediction tasks. 

\subsection*{Permutation-invariant electrolyte representation}\label{sec:perm_inv}
To achieve permutation invariance, we design an electrolyte representation for both property prediction and electrolyte design using a multi-head self-attention-based aggregation mechanism~\cite{vaswani2023attentionneed} (see Fig.~\ref{si_fig:model_architecture}c). 
The aggregation mechanism is adapted from~\citet{Zhang2024}'s work with two key modifications. 
First, before passing the molecular embedding through the attention blocks, it is scaled by the molar ratio to ensure that molecules with a ratio of zero do not contribute to the mixture embedding. 
Second, multi-head attention is incorporated to enhance the model’s expressiveness. 

More specifically, the permutation invariance is achieved with the following aggregation mechanism: 
\begin{gather}
    \bm{X} = [\bm{X_1}, \bm{X_2}, ...]^T,\\
    \bm{r} = [r_1, r_2, ...]^T, \\  
    \text{Aggr}(\bm{X}, \bm{r}) = \bm{r}^T \cdot \text{MultiHeadAttention}(\bm{r} \odot \bm{X})
\end{gather}
where $\bm{X}$ is the tensor that stores molecular embedding, and $\bm{r}$ is a vector with molar ratio information. 
\enquote{$\cdot$} is inner product and \enquote{$\odot$} is row-wise multiplication. 
The aggregation mechanism is permutationally invariant, as the inner product operation gathers information independent of the ordering of molecules. 

\subsection*{Empirical temperature and concentration dependence of conductivity}\label{sec:empirical_relation}
We incorporate prior domain knowledge into conductivity prediction through an empirical relation. 
The relation used in this work is described by the following equation:
\begin{gather}
    \sigma(T, c) = \frac{A}{\eta}c^{n_1}e^{-\frac{B \times c^{n_2} + D}{T - T_0}}
\label{eq:empirical_equation}
\end{gather}
where $A$, $n_1$, $B$, $n_2$, $D$ and $T_0$ are all learnable parameters from electrolyte embeddings but independent from temperature ($T$) and Li salt concentration ($c$, Li molar ratio). 
$\eta$ is the estimated viscosity of the solvent mixture based on predicted viscosities from molecule pretraining (see Supplementary Information). 
The inclusion of viscosity in the prediction of conductivity is inspired by the well-known Walden's rule~\cite{Walden1906}, which describes the inverse relation between conductivity and viscosity. 
The empirical relation in Eq.~\ref{eq:empirical_equation} is further validated using data from both the Advanced Electrolyte Model (AEM)~\cite{Zhu2024, Gering2006, Gering2017} and experiments~\cite{Ding2001}. 
As Fig.~\ref{si_fig:emp_relation} and Table~\ref{si_tab:empirical_relation} indicate, the chosen empirical relation fits well across various electrolyte systems, and the low loss from our predictive model further demonstrates the generality of this relation. 

\subsection*{Pretraining with computational data}\label{sec:comp_pretrain}
The computational dataset for pretraining the predictive model was randomly sampled within the chemical species of all solvent molecules in the experimental dataset, together with the electrolyte formulations in the experimental dataset. 
During the sampling, the maximum number of solvents was set to six, and LiPF$_{6}$ and LiFSI were chosen as the salts given their common usage in commercialized electrolytes. 
In total, there are 104,657 formulations with their MD simulated anion coordination ratio and ionic conductivities calculated by the Nernst-Einstein method and a method described by ~\citet{Mistry2023} in the computational dataset. Details of the MD simulations are provided in the Supporting Information. 

As briefly mentioned in the main text, the pretraining with computational data was performed to predict both conductivities---derived from the NE and Mistry's methods---as well as the anion ratio. Conductivities are predicted using an empirical relation, while the anion ratio is obtained directly through an MLP readout block. The readout block receives the electrolyte embedding, combined with temperature and concentration, as input and produces the corresponding prediction. During computational pretraining, the molecular embeddings obtained from molecular pretraining are kept frozen. Only the aggregation block used to generate the electrolyte embedding, along with the readout blocks for property prediction, remain trainable. 
\subsection*{Fine-tuning with experimental data}
Given the inaccuracy of MD-derived conductivity, we further fine-tune our predictive model with experimental conductivity values. The experimental data used for fine-tuning contains 10,407 entries with 62 types of solvents and 17 types of Li salts. The data were collected from online databases and public publications (about 50 publications in total). In combination with 104,657 anion ratio samples obtained from MD simulations, the model is fine-tuned to jointly predict experimental conductivity and computational anion ratio. As in computational pretraining, the molecular embeddings remain fixed, while the electrolyte embeddings are trainable during fine‑tuning.
More technical details of fine-tuning can be found in Supporting Information. 
\subsection*{Conditional generation}\label{sec:cond_diff}
The conditional diffusion model we utilize in this work is based on Denoising Diffusion Probabilistic Models (DDPM)~\cite{ho2020denoisingdiffusionprobabilisticmodels} to generates invariant
electrolyte embeddings given target conductivity and anion ratio. 
DDPMs are a class of generative models that synthesize data by learning to reverse a diffusion process, where noise is progressively added to data and then removed step-by-step. 
This approach has proven highly effective in generating high-quality, diverse samples across various domains. 
To realize conditional generation, additional labels are processed as embeddings besides time embeddings to modulate the noisy data during both diffusion and denoising steps (see Supporting Information for more details). 

In the next step, the generated electrolyte embeddings are decoded into a set of molecular embeddings using a Set Transformer–inspired architecture~\cite{Lee2019}, which preserves permutation invariance throughout the decoding process. 
Although not explored in this work, these molecular embeddings could potentially be further converted into 2D molecular graphs or SMILES strings by training an additional molecular decoder, enabling the discovery of new electrolyte molecules. 
For simplicity, we currently match the generated embeddings to existing molecules in our database. 
As a result, each electrolyte formulation can be represented as a \enquote{BoM} vector (see Supporting Information for technical details). 
Each value within the vector represents the corresponding molar ratio of the molecule at the position. 

To train a conditional diffusion model, we first used our predictive model to synthetically create 30,000 different electrolyte formulations with conductivity and anion ratio labels. 
The synthetic dataset offers a more balanced coverage of the electrolyte design space compared to the experimental dataset and provides improved accuracy in conductivity evaluation compared to the computational dataset.
All 62 solvents and 2 Li salts ($\text{LiPF}_6$ and LiFSI) from experiments are randomly sampled to compose different formulations. The molar ratios of solvents and salts sum up to 1 with fixed salt concentration (10 mol \%) and temperature (25\textdegree{}C). 
We select 2 Li salts as they are most commonly used in today's commercialized lithium-ion battery. 
As a result, our generative modeling primarily focuses on the solvent design space, which is typically the main target in electrolyte design. Details about the synthetic dataset are included in the Supporting Information. 

\subsection*{Classifier-guided diffusion}\label{sec:cgd}
To generate electrolyte formulations with base molecules and constrained molar ratios, we employ a classifier-guided diffusion (CGD) method~\cite{Cgd2021}. 
The essential idea is that our electrolyte decoder already generates the \enquote{BoM} vector, which can be directly used to assess whether the generated formulation contains certain base molecules. 
This allows us to construct a classifier function based on the decoder for any given base-molecule scenario without the need to retrain the classifier for each condition. 
As a result, this approach significantly speeds up the adaptation of our model to new cases compared to approaches like classifier-free guidance. 
The CGD method guides the sampling towards target base formulation constraint by computing the gradient of the classifier’s log-probability with respect to noisy data at time $t$ (see Supporting Information for more technical details). 
Consequently, the gradient-based sampling approach guides the generation process toward satisfying the target constraint. 

In order to better predict the \enquote{BoM} vector from noisy data during sampling, we retrain the decoder on noisy electrolyte embeddings for the CGD generation. 
These embeddings are generated as the noisy data obtained in the forward diffusion process. 
An additional hyperparameter, gradient scale, is multiplied to the gradient to control the degree of guidance (see Supporting Information for the usage and selection of gradient scale). 
We utilize the maximum success rate under different gradient scales for visualization in Fig.~\ref{fig:generation}f.

\subsection*{Performance evaluation for formulation generation}\label{sec:eval_metric}
To evaluate the performance of our generative process, we proposed three different evaluation metrics: (1) MAPE which evaluates the deviation of generated properties from the targets; (2) \enquote{formulation diversity} which calculates the average minimal pairwise difference between generated formulation; (3) \enquote{molecular diversity} which computes the entropy given the occurrence frequency of each molecule within the generated formulations. 
Together, these evaluation metrics offer insights into the accuracy of the generated formulations relative to the target, as well as their diversity in terms of molar ratio variation and molecular species composition. 
Mathematical expressions of all three metrics can be found in Supporting Information. 

All evaluations of the generative model’s performance are conducted on 10,240 generated electrolyte formulations for each case, produced in 40 batches with a batch size of 256. 
For evaluating the success rate of base formulation–constrained generation, we define a generated formulation as a successful generation if it satisfies the following criteria: (1) it meets the specified base formulation constraints of our target (e.g., the molar ratio of EC exceeds 20\%); and (2) the deviations in both conductivity and anion ratio of generated formulations are smaller than the corresponding standard deviations of those properties in the labeled dataset. 
The final success rate is calculated by dividing the number of generated formulations that meet these criteria by the total number of generated samples.
\subsection*{Conductivity measurement}\label{sec:conductivity_measurement}
All the experiments in this work were conducted by HEFEI FANGTIYA LLC. (\url{https://m.zhipin.com/companys/0507e6317c04b3f803x73tq8ElM~.html}). The solvents used to prepare liquid electrolytes were purchased from various vendors including Shinghwa Advanced Material Group Co., Ltd. (EC, PC, MA, DEC, EA, EP, DME), Aladdin Scientific (THF, DEM, METHF, EMP, DMM, EFA, MTFA), Xiamen Shouneng Technology Co., Ltd. (DOL, EGDEE), Guangzhou Tinci Materials Technology Co., Ltd. (FEC), Hi-Tech Spring (DMC), Hunan Zhonglan Network Technology Co., Ltd. (LiPF6), and Do-Fluoride New Materials Co., Ltd. (LiFSI). All materials were mixed at room temperature (25 \textdegree{}C), and the conductivities were measured using a conductivity meter (FE32-Standard, Mettler-Toledo Group, Switzerland). Each formulation was prepared and measured three times to ensure consistency and reliability. 

\subsection*{Raman spectroscopy}\label{sec:raman_spectroscopy}
Raman measurements were conducted using a DXR2 Raman microscope. A 532 nm laser was used for most samples, while a 785 nm laser was applied for electrolytes that potentially exhibited fluorescence background. The spectral range covered wavenumbers from 100 to 4250 cm$^{-1}$. All the experiments were performed by Hefei Fangtiya New Material Technology Co., Ltd.. From Figure~\ref{fig:exp_validation}b, we can see that the peak positions of the single solvent-LiFSI electrolyte have the trend of DMC $>$ DEC $>$ MA $>$ EA $>$ THF $>$ EC $>$ PC, which aligns well with Ref.~\cite{su2019solvating} and demonstrates the validity of our Raman spectroscopy measurement.
\backmatter

\bmhead{Supplementary information}
Additional explanations and details regarding the datasets, model architectures, training and evaluation approaches, and results can be found in the supplementary information.



\section*{Declarations}
\subsection*{Conflicts of interest}
ByteDance Inc. holds intellectual property rights pertinent to the research presented herein. 
\subsection*{Author contribution}
Conceptualization: Zhenze.Y., S.G. and W.Y.; Methodology: Zhenze.Y., S.G., Y.W., W.Y., X.H., Z.Z. and H.L.; Investigation: Zhenze.Y. Y.W., X.H., Z.Z., H.L., Z.M., T.Z., S.L., Z.P., Z.W., Zhiao.Y., S.G., W.Y.; Supervision: S.G. and W.Y.; Writing: Zhenze.Y., S.G. and W.Y.
\subsection*{Data availability}
The computational and experimental literature datasets, together with checkpoints for both predictive and generative models are available on HuggingFace (\url{https://huggingface.co/ByteDance-Seed/bamboo_mixer}).
\subsection*{Code availability}
The source codes of both predictive and generative workflow to reproduce the results from this work are public on Github (\url{https://github.com/ByteDance-Seed/bamboo_mixer}). 


\begin{appendices}




\end{appendices}


\clearpage
\begin{figure}[htbp!]
\centering
\includegraphics[width=0.99\linewidth]{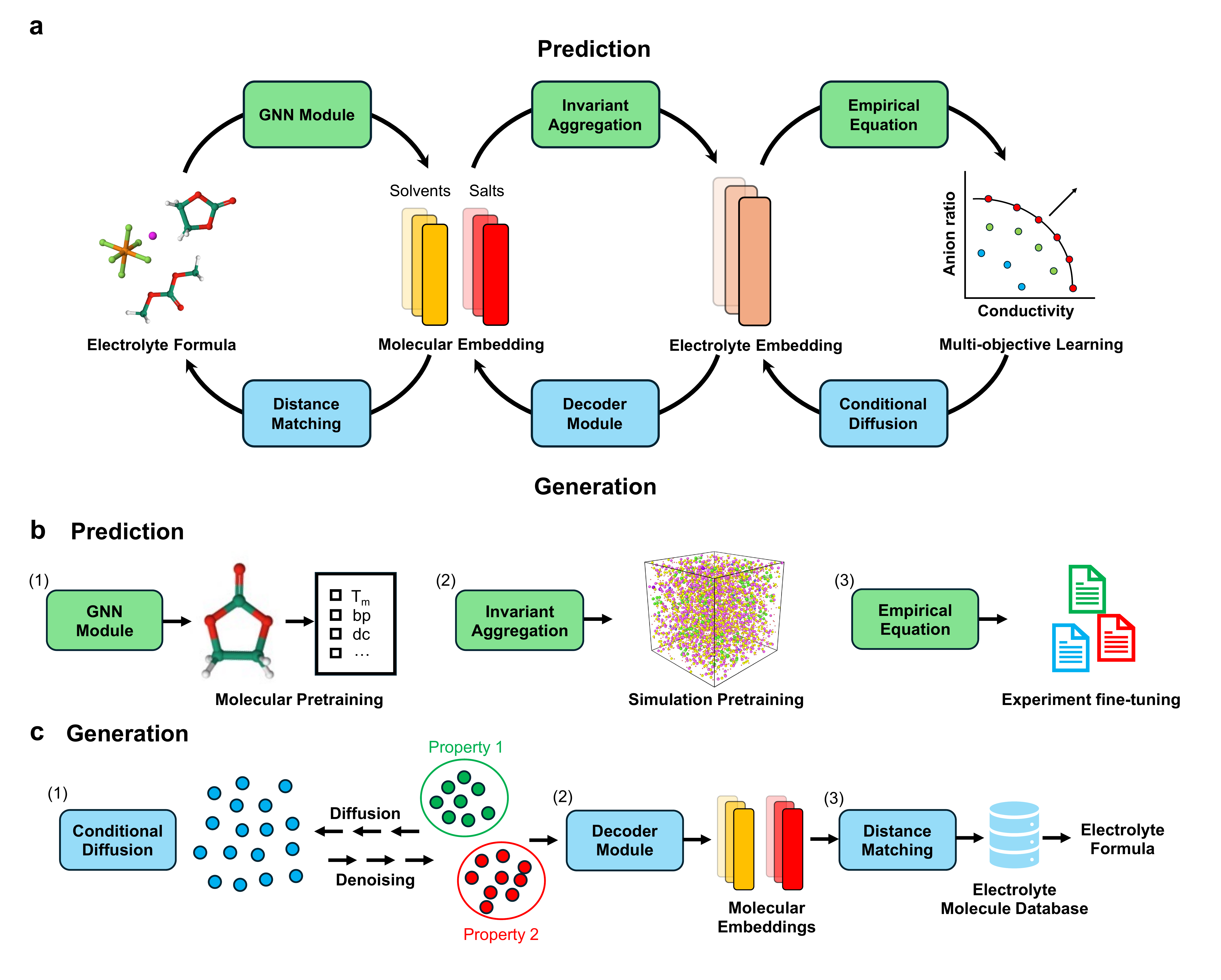}
\caption{\textbf{A predictive and generative electrolyte design workflow reported in this work.} \textbf{a} Forward (prediction) and inverse (generation) processes of electrolyte formulation are designed as three-stage workflows, using molecular embeddings (representations of individual component molecules in an electrolyte formulation) and electrolyte embeddings (permutation-invariant representations of entire electrolyte formulation). \textbf{b} Three stages of predictive model for conductivity and anion ratio predictions: (1) A GNN model is trained on single-molecule dataset on multi-property prediction, generating universal molecular embeddings. (2) MD data of around 100,000 different electrolyte formulations are utilized to further construct an informative electrolyte embedding from molecular embeddings. (3) An empirical relation is integrated into the model architecture and fine-tuned with 10,000+ experimental literature conductivity data points. \textbf{c} Three stages of generative model given property conditions: (1) A conditional diffusion model generates electrolyte embeddings based on specified properties. (2) The generated electrolyte embeddings are converted back to molecular embeddings with a decoder. (3) Finally, molecular embeddings are matched with our chemical database to obtain the electrolyte formulation. }\label{fig:overview}
\end{figure}

\clearpage
\begin{figure}[htbp!]
\centering
\includegraphics[width=0.99\linewidth]{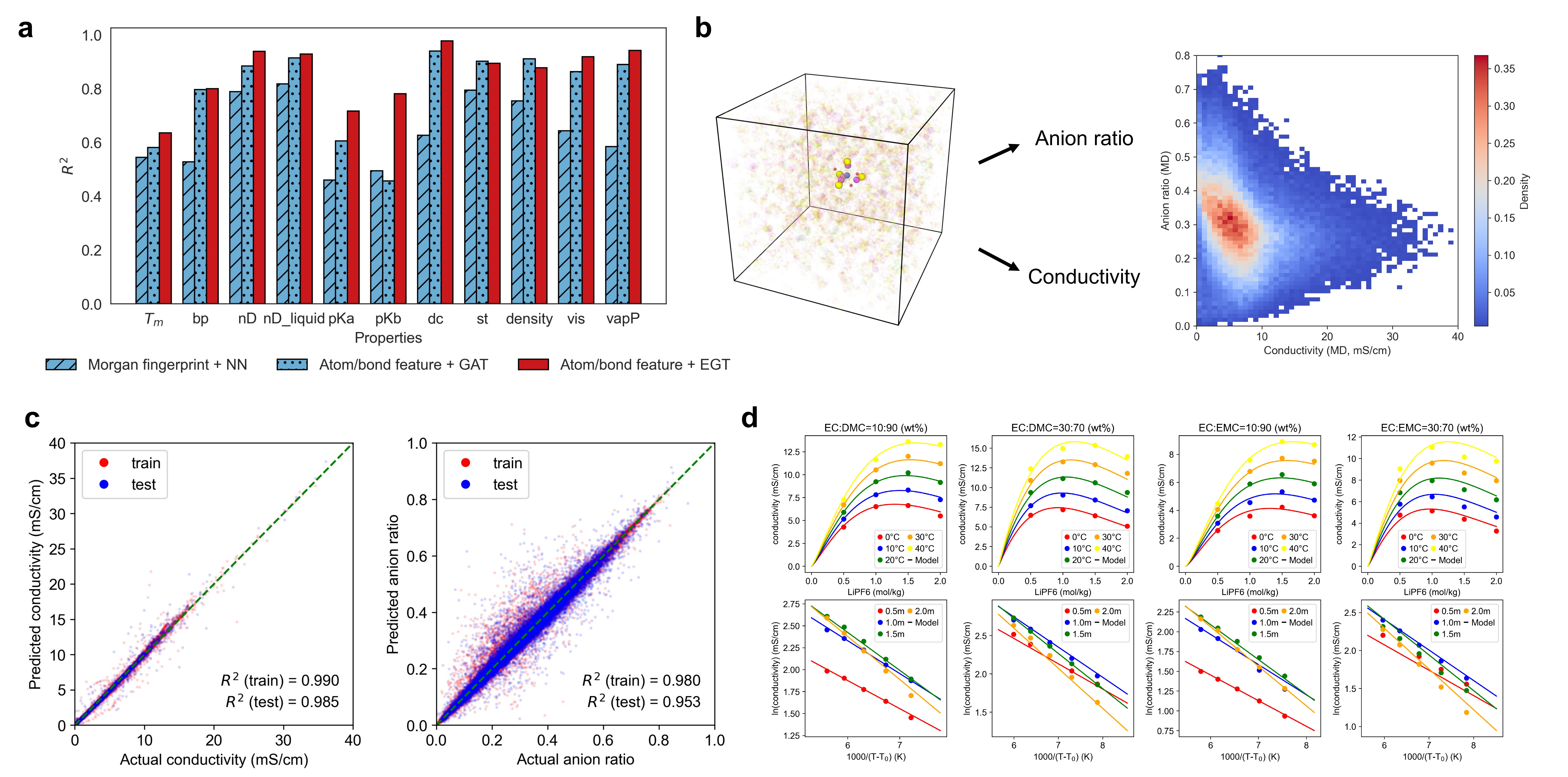}
\caption{\textbf{Prediction performance} \textbf{a} Comparison of model performance across various molecular properties during molecular pretraining. Our GNN model utilizes atom and bond features as input and is based on an EGT model (red bar, \enquote{Atom/bond feature + EGT}). Other two models are considered as baseline results (blue bars, \enquote{Morgan fingerprint + NN} and \enquote{Atom/bond feature + GAT}). \textbf{b} Anion ratio and ionic conductivity obtained from MD simulations using an OPLS force field. The density of data below 0.005 is omitted in the figure. Here, conductivity calculated using Mistry's method is used given that it aligns generally better with experiments (Fig.~\ref{si_fig:cond_compare}). \textbf{c} Predictions of our model versus ground truth after experimental fine-tuning for both anion ratio and conductivity. \textbf{d} Example predictions of temperature and concentration dependence of conductivity across various electrolyte systems with empirical equation. $T_0$ is a learnable temperature parameters in the empirical equation, which is generally related to glass transition temperature of electrolytes. }\label{fig:prediction}
\end{figure}

\clearpage
\begin{figure}[htbp!]
\centering
\includegraphics[width=0.99\linewidth]{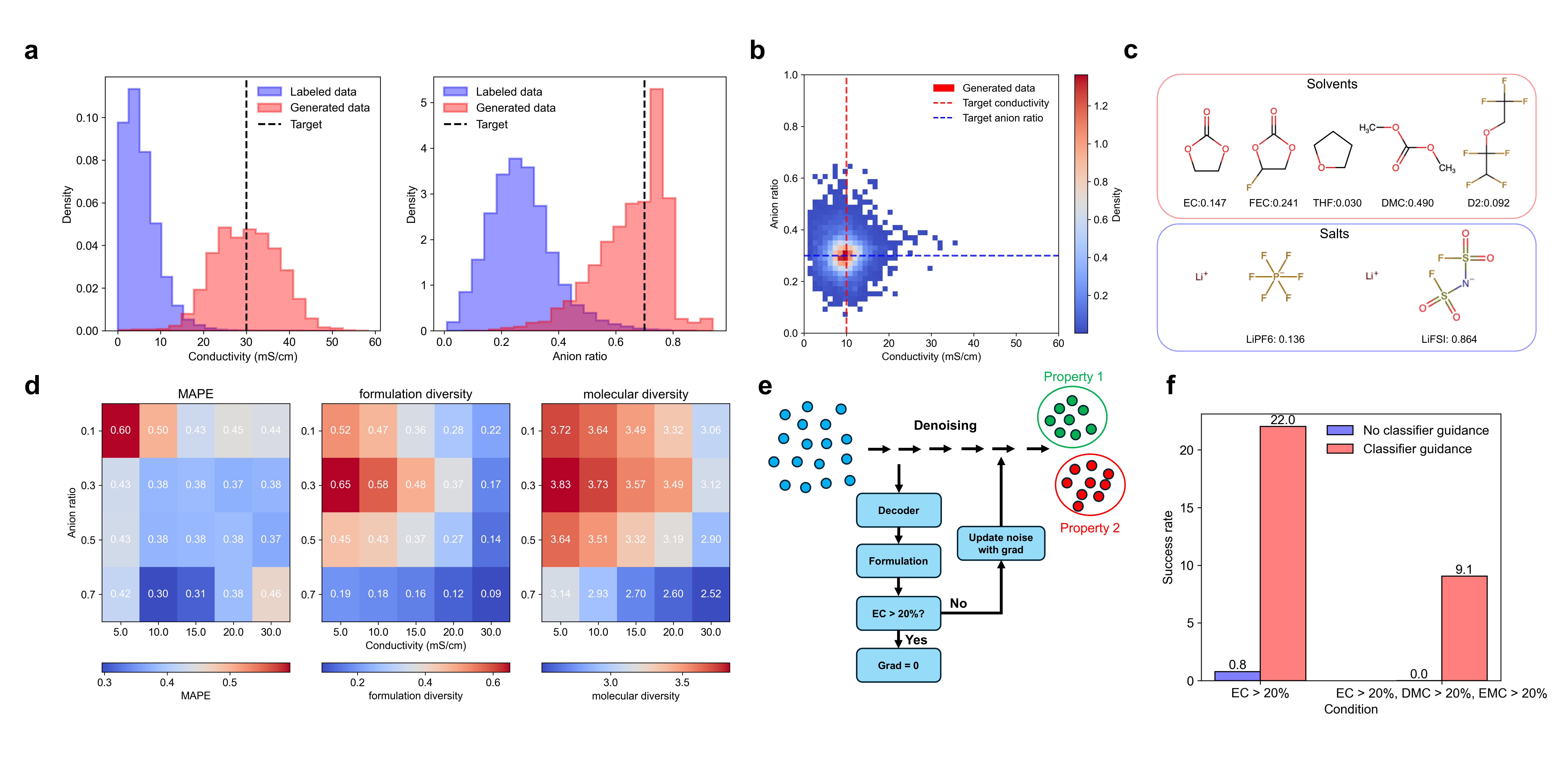}
\caption{\textbf{Generation performances} \textbf{a} Extrapolation of target properties using generative model. To extrapolate conductivity, the anion ratio is fixed at the dataset’s mean value. Similarly, to extrapolate the anion ratio, conductivity is set to its mean. \textbf{b} Example distribution of generated electrolyte formulations given target conductivity (10.0 mS/cm) and anion ratio (0.3). More results of different conditions can be found in Fig.~\ref{si_fig:cond_generation}. \textbf{c} One example of generated electrolyte formulation from panel Fig.~\ref{fig:generation}b. More examples are listed in Fig.~\ref{si_fig:example_formulation_5.0} - Fig.~\ref{si_fig:example_formulation_30.0}. \textbf{d} Evaluation of generation using three metrics including MAPE and two diversity scores. \textbf{e} Schematic of classifier-guided conditional diffusion to ensure that the generated formulation satisfies the base formulation constraints. \textbf{f} Performance comparison between conditional generation and classifier-guided generation in satisfying the base formulation constraints. 
}\label{fig:generation}
\end{figure}

\clearpage
\begin{figure}[htbp!]
\centering
\includegraphics[width=0.99\linewidth]{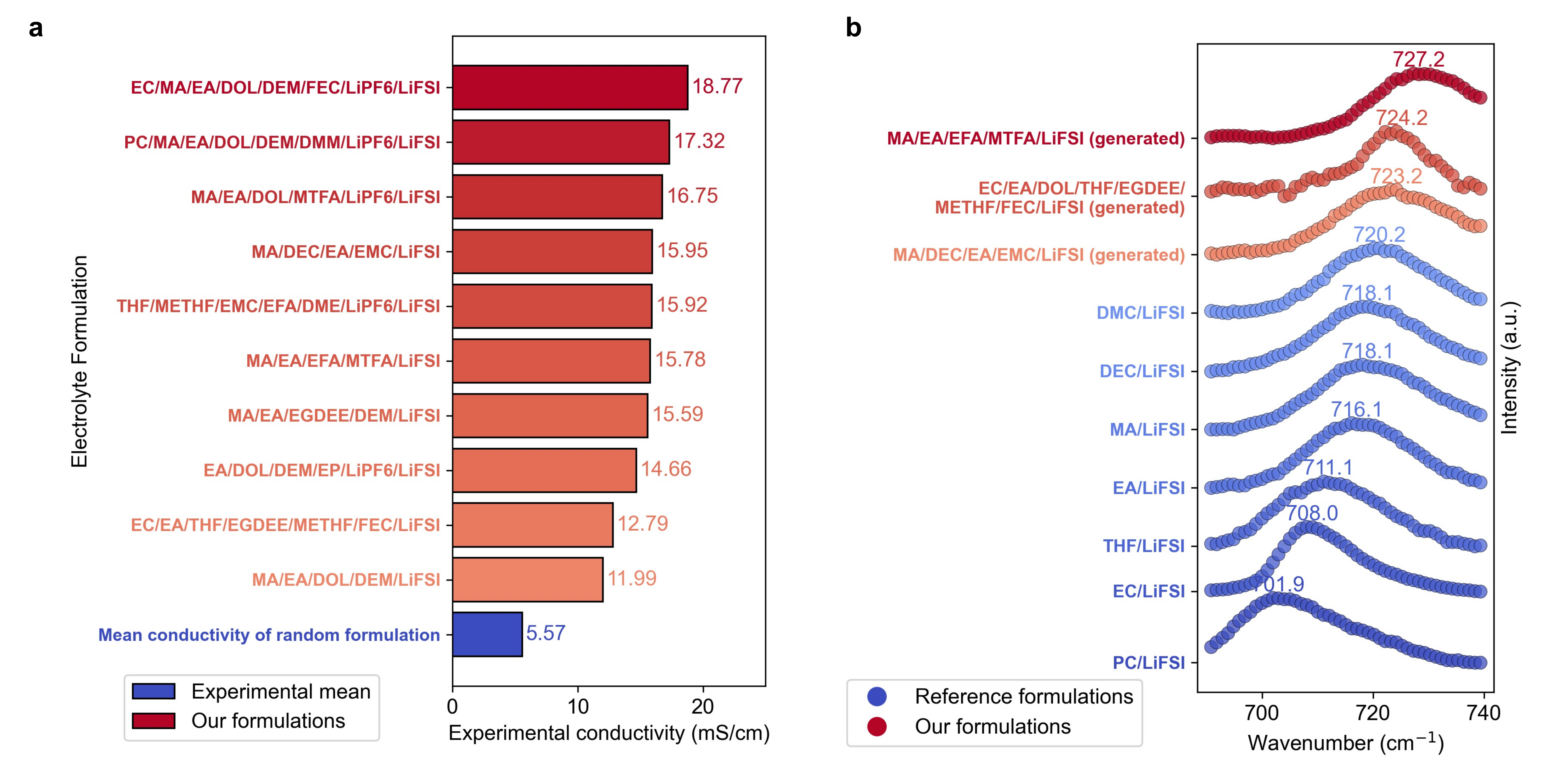}
\caption{\textbf{Experimental validation} \textbf{a} Conductivities of generated formulations measured from experiments (target conductivity in conditional generation = 15 or 20 mS/cm). \textbf{b} Raman spectrum of FSI$^{-}$ comparing generated formulations with single-solvent systems. The detailed molar ratios of each solvent component in the formulations are omitted from the figure and can be found in Table~\ref{si_tab:exp_formulations}.
}\label{fig:exp_validation}
\end{figure}

\newpage
\appendix

\renewcommand{\theHfigure}{S\arabic{figure}}
\renewcommand{\theHtable}{S\arabic{table}}

\section*{Supporting Information}
\addcontentsline{toc}{section}{Supporting Information}
\tableofcontents
\newpage
\setcounter{figure}{0}
\setcounter{table}{0}
\setcounter{equation}{0}
\renewcommand{\thetable}{S\arabic{table}}
\renewcommand{\thefigure}{S\arabic{figure}}
\section{Data representation}\label{sec_SI_data}
The detailed data representations, along with their corresponding notations and shapes used in this work, are provided in Table~\ref{si_tab:data_repr}.

\subsection{Molecular representation}
For molecular representation, we use a 2D graph to model each individual molecule, where (mapped) SMILES strings are utilized to extract atomic and bond features. The atomic features include: element type, ring connectivity, minimum ring size, and formal charge. The bond features are bond order and whether the bond belongs to a ring. Our GNN model takes these features as input and create a 1D vector for molecular embedding of dimension 64 (denoted as \enquote{$\bm{h}_m$} in the following context). For performance benchmarking, we also used a 512-bit Morgan fingerprint generated by RDKit~\citeSI{rdkit} as the molecular descriptor to train a baseline model.

\subsection{Electrolyte representation}
The electrolyte formulation is represented by either an electrolyte embedding (denoted as \enquote{$\bm{h}_{e}$} in the following text) or a \enquote{Bag-of-Molecules} (BoM) vector (denoted as \enquote{$\bm{v}_{\text{BoM}}$} in the following text) when dealing with fixed chemical space in this work. The electrolyte embedding is a 384-dimensional vector derived from molecular embeddings using permutation-invariant aggregation operations. Notably, individual molecular contributions within the formulation can not be explicitly separated from the electrolyte embedding. The \enquote{BoM} vector follows a similar concept of \enquote{Bag-of-Words} from NLP. It has a dimensionality equal to the number of distinct molecules in our molecular \enquote{vocabulary}, where each position corresponds to a specific molecule, and the value represents its molar ratio (in NLP, it is the occurrence frequency of each word). For instance, in our generation scenario, the vector is structured such that solvents occupy the first 62 dimensions, while the last 2 dimensions correspond to salts. In addition, molar ratios are handled separately for solvents and salts—specifically, the sum of the molar ratios for solvents is 1, and the same holds for salts.

\section{Molecular pretraining}\label{sec_SI_pretrain}
\subsection{Single-molecule dataset}\label{sec_SI_mono}
The single-molecule dataset contains molecular property information gathered from 11 public datasets as shown in Table~\ref{si_tab:mono_dataset}. The SMILES of molecules are collected from these databases or determined based on their names using CIRpy~\citeSI{Peach2018} which is a python interface for resolving chemical identifiers such as names, CAS registry numbers and SMILES strings. We then screened the data using a combination of following principles to remove incorrect, inconsistent or ambiguous entries: 
\begin{enumerate}
    \item Remove molecules with \enquote{bad} SMILES. A SMILES is a good SMILES if it satisfies: 
    \begin{itemize}
        \item \enquote{.} does not appear in SMILES (individual molecule).
        \item elements covered by H, C, N, O, F, P, S, Cl, Br, I.
        \item formal charge for F, Cl, Br, I must be negative.
        \item halogen connectivity must be 1.
        \item hybridization in $s, sp, sp^2, sp^3$.
    \end{itemize}
    \item Removing data from different sources that exhibit high inconsistencies.
    \begin{itemize}
        \item large standard deviation for same properties.
        \item different temperature dependence. 
    \end{itemize}
    \item Removing ambiguous data such as \enquote{$T_m > 300K$}.
\end{enumerate}
After data preprocessing and screening, the final dataset consists of 241,414 entries with 11 molecular properties: $\bm{T_m}$: melting point, $\bm{T_b}$: boiling point, $\bm{n_D}$/$\bm{n_D^{\textbf{liquid}}}$: (liquid) refractive index, $\bm{pK_a}$: acid dissociation constant, $\bm{pK_b}$: base dissociation constant, $\bm{\varepsilon}$: dielectric constant, $\bm{\gamma_s}$: surface tension, $\bm{\rho}$: mass density, $\bm{\eta}$: viscosity and $\bm{P_{\text{vap}}}$: vapor pressure. Details regarding the final dataset, along with the corresponding data sources and the number of entries for each property label, are provided in Table~\ref{si_tab:mono_dataset}.

\subsection{Multi-task learning}\label{si_sec:multi-task learning}
Multi-task learning is employed to generate a universal molecular fingerprint for downstream formulation-level learning. The shared molecular fingerprint is then passed through separate readout blocks to predict different molecular properties. Of the 11 properties, density, refractive index, and dielectric constant are temperature-dependent but only contains data at a single temperature per molecule, whereas viscosity, surface tension, and vapor pressure are characterized over multiple temperatures. For the single‐temperature properties, we simply concatenate the temperature with the molecular embedding for prediction. In contrast, to accurately capture the temperature dependence of viscosity, surface tension, and vapor pressure and to align our outputs with empirical curves, we incorporate empirical equation blocks for these properties and use MLPs to learn the corresponding parameters within the empirical equation. More specifically, the surface tension generally decreases linearly as temperature increases (Eötvös rule)~\citeSI{Mitra1954}, the relation between viscosity and temperature observes Vogel–Fulcher–Tammann (VFT) equation~\citeSI{ngai2011relaxation} and vapor pressure follows the Clausius-Clapeyron (CC) equation~\citeSI{Brown1951}. The corresponding empirical relations are simplified as follows:
\begin{gather}
    A^{\gamma_s} = \text{MLP}(\bm{h}_m, \text{act=softplus}) \\
    B^{\gamma_s} = \text{MLP}(\bm{h}_m, \text{act=softplus}) \\
    \gamma_s = -A^{\gamma_s} \times T + B^{\gamma_s}
\label{si_eq:1}
\end{gather}
\begin{gather}
    \text{ln}(\eta_0) = \text{MLP}(\bm{h}_m, \text{act=softplus}) \\
    A^{\eta} = \text{MLP}(\bm{h}_m, \text{act=softplus}) \\
    T_0^{\eta} = \text{MLP}(\bm{h}_m, \text{act=softplus}) \\
    \text{ln}(\eta) = \text{ln}(\eta_0) + \frac{A^{\eta}}{T-T_0^{\eta}}
\label{si_eq:2}
\end{gather}
\begin{gather}
    A^{P_{\text{vap}}} = \text{MLP}(\bm{h}_m, \text{act=softplus}) \\
    B^{P_{\text{vap}}} = \text{MLP}(\bm{h}_m, \text{act=softplus}) \\
    \text{ln}(P_{\text{vap}}) = A^{P_{\text{vap}}} - \frac{B^{P_{\text{vap}}}}{T}
\label{si_eq:3}
\end{gather}
where $T$ is the temperature, $A$s and $B$s are learnable parameters (all positive) based on the molecular embedding ($\bm{h}_m$). The superscript denotes the relevant property, while the subscript indicates whether it is the reference state of that property. The \enquote{softplus} activation function is used for all the MLP blocks listed above to ensure positive learnable parameters and prevent gradient vanishing issues. For those properties without empirical relation, we use \enquote{sigmoid} function for their readout layers. 

\begin{figure}[htbp!]
\centering
\includegraphics[width=0.99\linewidth]{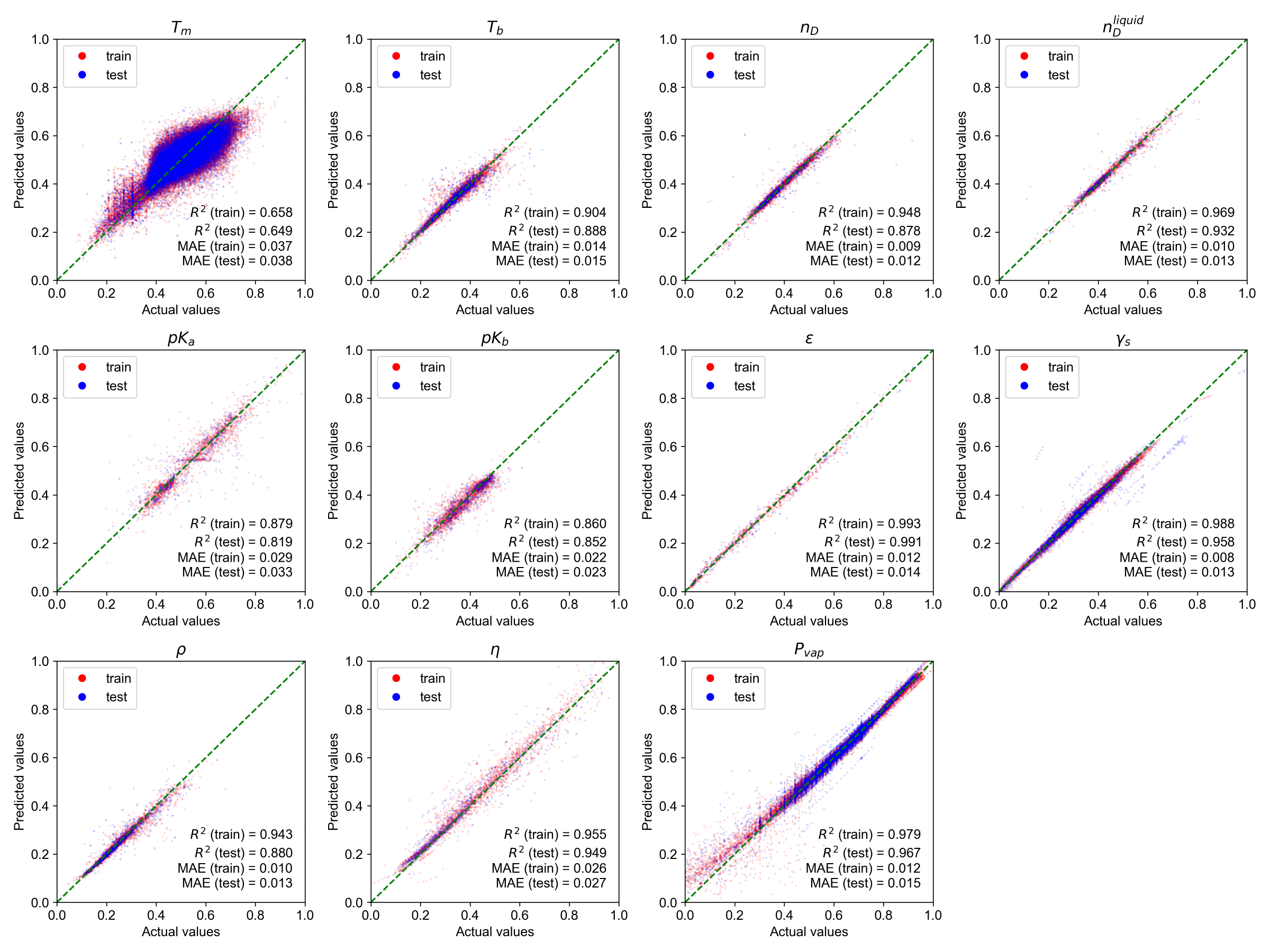}
\caption{\textbf{Predicted molecular properties from molecular pretraining.} The parity plots comparing molecular properties predicted from our GNN model and ground truth from experiments. }\label{si_fig:pretrain_scatter}
\end{figure}

All the molecular properties are normalized to values between 0 and 1 based on their minimum and maximum values within the dataset. The total loss of multi-task learning is simply a weighted sum of Mean Squared Error (MSE) loss of each individual property. Here the weights are set to be 1 for all properties. The total loss is therefore as follow:
\begin{gather}
     \mathcal{L}_{\text{tot}} = \mathcal{L}_{T_m} + \mathcal{L}_{T_b} + \mathcal{L}_{n_D} + \mathcal{L}_{n_D^\text{liquid}} + \notag \\ \mathcal{L}_{pK_a} + \mathcal{L}_{pK_b} + \mathcal{L}_{\varepsilon} + \mathcal{L}_{\gamma_s} + \mathcal{L}_{\rho} + \mathcal{L}_{\eta} + \mathcal{L}_{P_{\text{vap}}}
\end{gather}
The whole dataset is split into train (70\%) and test (30\%) sets for training and evaluation. As mentioned in the main text, we employ the graph blocks of the ByteFF model~\citeSI{Zheng2025} to process atomic and bond information, generating a molecular embedding or fingerprint. The atom and edge features are first transformed into node and edge embeddings through an MLP block with two hidden layers of 32 neurons and GELU activation. These embeddings are then processed by three EGT layers, with a hidden dimension of 32 and 4 attention heads. The dimensions of both node and edge embeddings generated by the graph block are set to 32, resulting in a molecular embedding of dimension 64. The readout MLP contain three hidden layers with 256, 128 and 16 neurons respectively. MAE is used as the loss function for each property. When a label is unavailable for a specific molecule, a masking mechanism is applied to exclude the corresponding loss from the total loss function and the loss of each properties is averaged based on the number of labels. The model is trained with a batch size of 512, and early stopping with a patience of 50 epochs is applied to monitor and optimize training. A comparison between predicted properties and ground truth data is presented in Fig.~\ref{si_fig:pretrain_scatter}.

\subsection{Comparison with baseline models}
To benchmark the performance of multi-task learning and ensure the effectiveness of the molecular embeddings, we developed and trained two additional baseline models: one is a simple fully connected neural network using the Morgan fingerprint as input features (labeled as \enquote{Morgan fingerprint + NN} in Table~\ref{si_tab:pretrain_comparison}), and the other is a GNN model in which the EGT layers are replaced with GAT layers (labeled as \enquote{Atom/bond feature + GAT} in Table~\ref{si_tab:pretrain_comparison}). We selected a 512-bit Morgan fingerprint as input and passed it through an MLP block with two hidden layers (256 and 128 neurons) to match the dimensionality of the molecular fingerprint generated by the EGT model. In terms of GAT layers, we selected the same hyperparameters (hidden dimension = 32, number of attention heads = 4, attention channel = 8, number of layers = 3) as the EGT layers to ensure a fair comparison. The generated molecular embeddings from baseline models are passed through the same MLP and empirical equation blocks to predict molecular properties as described in Section~\ref{si_sec:multi-task learning}. All three models have approximately 800,000 trainable parameters and are trained using the same strategy for performance comparison. 

The performances of these models, evaluated using $R^2$ and MAE, are displayed in Table~\ref{si_tab:pretrain_comparison}. As the table shows, the EGT-based model outperforms the two baseline models in all property prediction tasks.
\section{Electrolyte property prediction}
\subsection{Computational dataset}
To generate the computational dataset, MD simulations were conducted using LAMMPS~\citeSI{Thompson2022}, with the OPLS-AA force field~\citeSI{Jorgensen1996, lSoetens1998, Sambasivarao2009}. The atomic charge was set as 0.8 times the CHELPG partial charges~\citeSI{Breneman1990} calculated by the GPU4PySCF software~\citeSI{Wu2025}. 
The system size and simulation length of the MD simulations were set to be 10,000 atoms and 10 ns with the time step of 2 fs, respectively. 
To focus on the effect of coverage of chemical space instead of temperature, MD simulations are conducted at constant temperature 298 K (25 \textdegree{}C).
Anion ratio was calculated by counting the number of molecules and ions within 2.8 \AA\ of Li ions and dividing the number of anions by the total number of counted species. 

In this work, two approaches were used to calculate ionic conductivities from MD simulations, the Nernst-Einstein method and a method described by ~\citetSI{Mistry2023} that built on Stefan-Maxwell diffusivities. The Nernst-Einstein equation is described below: 
\begin{equation}
    \sigma_{NE} = \cfrac{e^2}{Vk_B T} (N_+z^2_+ \bar{D}_++N_-z^2_- \bar{D}_-),
\label{eq:NE}
\end{equation}
where $e$ is the elementary charge, $k_B$ is the Boltzmann constant, $V$ and $T$ are volume and temperature of the simulated system, and $\bar{D}_{\pm}$, $z_{\pm}$, and $N_{\pm}$ are self-diffusion coefficients, valence charges, and number of the positive and negative ions, respectively. The NE relation neglects ion pairing and the correlated motion of molecules in MD simulations, causing it to overestimate ionic conductivity. In contrast, Mistry’s method~\citeSI{Mistry2023} incorporates the relative motion of neutral molecules, cations, and anions. Consequently, Mistry’s conductivity estimates for typical liquid electrolytes are consistently lower than those from the NE relation, yet they converge to NE values at extremely low salt concentrations.

\subsection{Experimental dataset}
The experimental dataset of electrolyte conductivity is obtained by gathering data from public datasets and academic publications. There are two public datasets we collect data from: (1) Polymer Electrolyte Dataset~\citeSI{Bradford2023}; (2) CALiSol-23 Dataset~\citeSI{Blasio2024}. For the Polymer Electrolyte Dataset, we first filtered out all polymer electrolytes based on their SMILES representation, as polymer SMILES contain the symbol \enquote{$\ast$}. For the remaining data, we manually reviewed the source publications, examining each paper individually to remove inconsistent data (e.g. solid electrolyte, duplicate but inconsistent entries, ...) and correct any errors (e.g. unnormalized molar/weight ratio, incorrect SMILES, ...). We also removed the conductivity at extremely low temperature (below phase transition temperature) and concentration (salt molar ratio $<$ 0.0001). For CALiSol-23 Dataset, a similar data processing approach was applied. In addition to these two public datasets, we incorporate conductivity data from 21 more recent studies, with a focus on interfacial stability, particularly in fluorinated electrolyte systems~\citeSI{Sasaki2005,Yang2024,Yu2022,Amanchukwu2020,Huang2024,Chen2022,Cui2024a,Xu2024,Chen2024a,Liang2025,Cui2024b,Fan2024,Zhao2023,Tan2024,Wu2024a,Emilsson2025,Wu2024b, Nambu2015, Wang2016, Dave2022, Piao2020}.

During the data preprocessing, all ratios were converted to molar ratios. For data reported in weight ratio, it is straightforward to use molar weight for conversion. For data reported in volume ratios, we used density information if available from the paper, to convert them to molar ratios. For cases where density was not provided, we either searched for density values or utilized predicted densities from molecular pretraining for individual solvent molecules to estimate the density of solvent mixtures. We assume ideal mixing for estimating volumes of mixtures, which may introduce some errors in data processing. However, this approach is the best solution we can think of given the available information. As a result, the density of a solvent mixture is essentially the volume‑weighted average of the densities of its individual components. In addition, the addition of solid salts can also lead to volume changes. To account for this, we adopt an empirical observation to estimate the molar ratio of both solvents and salts: preparing an $xM$ electrolyte solution requires adding approximately $x(1 + 0.05x)$ moles of salt per liter of solvent.

As a result, we curated an experimental dataset of electrolyte conductivity consisting of 10,407 entries, including 62 types of solvents and 17 Li salts. This dataset covers a wide array of elements (H, Li, B, C, N, O, F, P, S, Cl) and chemical species (carbonates, ethers, nitriles, fluorinated solvents, phosphates, ...), which can provide comprehensive conductivity information for electrolyte design.

\subsection{Pretraining with computational data}
In the pretraining stage using computational data, the multi-task loss is described as:
\begin{gather}
         \mathcal{L}_{\text{tot}} = \mathcal{L}_{\text{conductivity(NE)}} +   \mathcal{L}_{\text{conductivity(mistry)}} +
         \mathcal{L}_{\text{anion ratio}} 
\end{gather}
Both conductivities are log-normalized based on the maximum and minimum conductivity values of experimental dataset. Anion ratio is inherently between 0 and 1. The batch size is set to 512. To predict the anion ratio, temperature and concentration are concatenated to the end of the electrolyte embedding. Both temperature and concentration are each repeated eight times before concatenation. As a result, the final feature vector has a dimension of 400. The readout block for anion ratio prediction is a simple MLP consisting of three hidden layers, followed by a sigmoid activation function. The hidden layers contain 512, 128, and 16 neurons, respectively. In addition, the empirical relation block is also used for predictions of conductivities calculated from MD results. 
\subsection{Fine-tuning with experimental data}
Experimental fine-tuning follows a similar multi-task learning approach as molecular and computational pretraining. The dataset comprises over 10,000 experimental conductivity measurements and more than 100,000 computational anion ratio samples, which are randomly split into training and test sets. Fine-tuning was initialized from the checkpoints obtained during computational pretraining, using conductivity predicted by Mistry’s method as the starting point for experimental conductivity prediction. We freeze the molecular embedding and allow updates of the electrolyte embedding during the fine-tuning. The MAE losses for conductivity and anion ratio are averaged within each batch based on the number of samples with corresponding labels. The total loss is the sum of MAE losses of conductivity and anion ratios:
\begin{gather}
         \mathcal{L}_{\text{tot}} = \mathcal{L}_{\text{conductivity}} + \mathcal{L}_{\text{anion ratio}} 
\end{gather}

To perform a generalization test of the model as shown in Table~\ref{si_tab:generalization}, the training and test sets are split based on whether the electrolyte formulations contain solvents with specific elements such as F, S, or P for experimental fine-tuning. We aim to evaluate whether computational pretraining can enhance the model’s ability to generalize to molecules not observed in experimental data. Therefore, the corresponding solvents are kept in computational pretraining. For generalization test on viscosity, we modified the empirical equation (with or without $\eta$ in the Eq.~\ref{si_eq:emp_relation}) during both computational pretraining and experimental fine-tuning to keep the comparison consistent. 
\section{Physics-informed architectures}\label{sec_SI_physics}
\subsection{Permutation invariance}
\begin{figure}[htbp!]
\centering
\includegraphics[width=\linewidth]{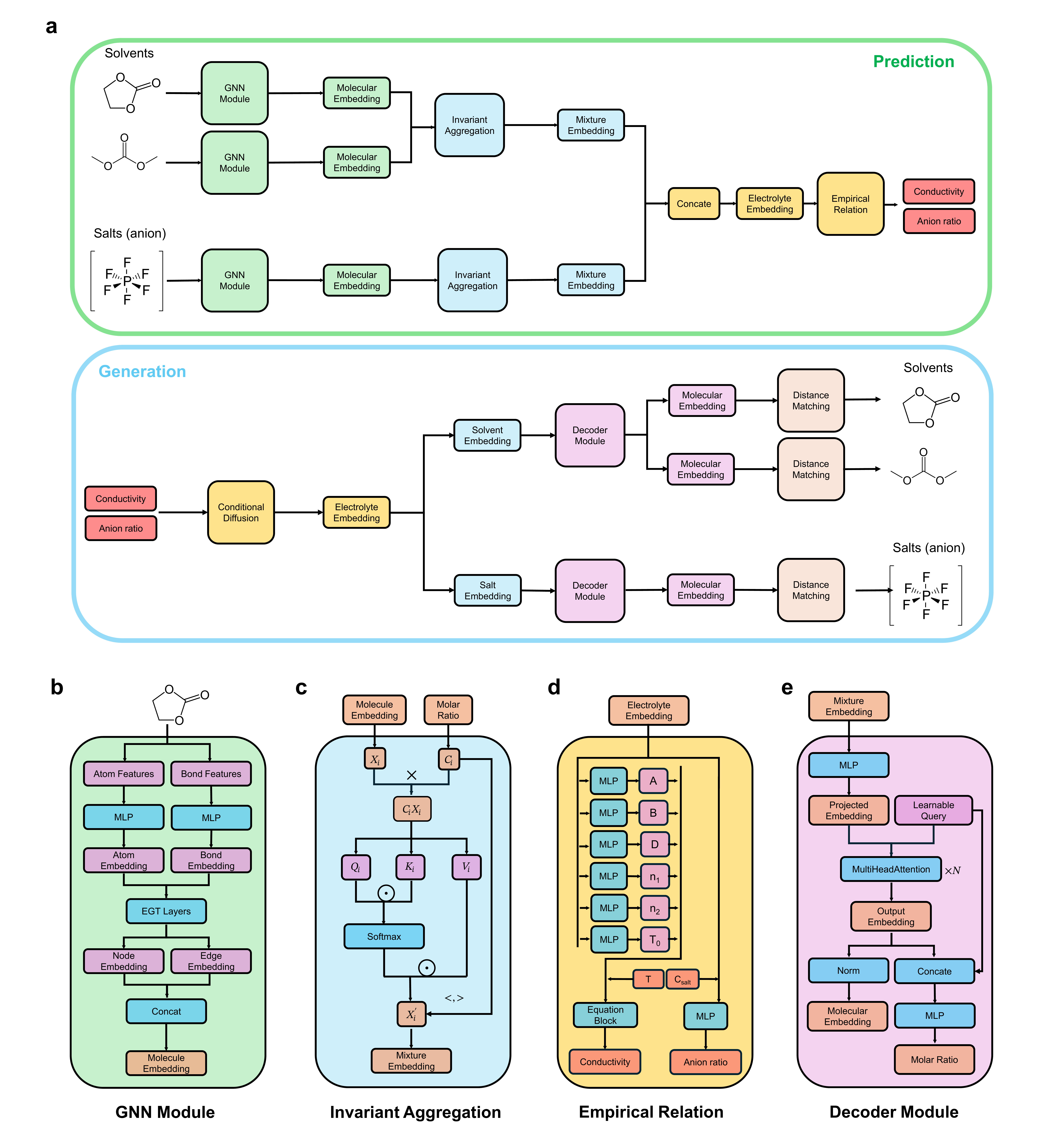}
\caption{\textbf{Detailed model architectures within the workflow reported in this work.} \textbf{a} Overall workflow of both predictive and generative process. Solvents and Li salts are considered separately in the workflow. \textbf{b} Module architecture of GNN model which takes SMILES of a single molecule as input and generates a universal molecular embedding by multi-task learning on 11 molecular properties. \textbf{c}. A self-attention-based aggregation block to merge multiple molecular embeddings into a mixture embedding and ensure permutation invariance. \enquote{$\times$} is row-wise multiplication, \enquote{$\odot$} stands for matrix multiplication and \enquote{$\langle, \rangle$} represents inner product. \textbf{d} The empirical relation block for conductivity and anion ratio prediction. For conductivity, there are six empirical learnable parameters (except viscosity) based on the electrolyte embedding, while anion ratio is predicted directly from the electrolyte embedding concatenated with temperature and concentration using a readout MLP layer. Activation layers are omitted from the schematic. \textbf{e} The decoder module which recovers molecular embeddings from mixture embeddings for both solvents and salts, which is further matched to molecules in our electrolyte molecule database. } 
\label{si_fig:model_architecture}
\end{figure}

Permutation invariance is realized using a self-attention mechanism with molar ratio-based scaling as described in Section~\ref{sec:perm_inv}. The multi-head attention uses classical query ($Q$), key ($K$) and value ($V$) to compute the attention output (see Fig.~\ref{si_fig:model_architecture}c):
\begin{gather}
    Q = W_Q(\bm{r} \odot \bm{X}) + B_Q\\
    K = W_K(\bm{r} \odot \bm{X}) + B_K\\
    V = W_V(\bm{r} \odot \bm{X}) + B_V\\
    \bm{X'} = \text{softmax}(\frac{QK^T}{\sqrt{d_K}})V
\end{gather}
where $W$ and $B$ are learnable weights and biases in the self-attention mechanism.  \enquote{$\odot$} stands for row-wise multiplication and $\bm{X}$ is the molecular embedding tensor and $d_k$ is the dimensionality of $K$. To maintain a consistent shape for $\bm{X}$ in the attention mechanism, we pad the molecular embeddings of both solvents and salts to a maximum of 12 molecules (12 molecules for solvents and 12 molecules for salts). The attention output is then aggregated using an inner product with molar ratio vector ($\bm{r}$) to ensure permutation invariance:
\begin{gather}
    \bm{h_{m}} = \bm{r^T} \cdot \bm{X'}
\end{gather}
\enquote{$\cdot$} is the inner product. We separate the molecular embeddings of solvents and salts, processing them independently through the aggregation block. Since salts are typically fewer in number than solvents in the electrolytes, we use 2 attention heads for salts and 4 for solvents. All the hidden dimensions are selected as 64 for each attention head. The numbers of attention head are 4 and 2 for solvents and salts respectively. As a result, the electrolyte embedding has a dimension of $64 \times (4 + 2) = 384$. 
\subsection{Empirical relations}\label{si_sec:empirical_relation}
The empirical equation is integrated into our predictive model to enable physics-informed conductivity prediction. We here discuss about the derivation of empirical relation. 

In dilute solutions, the conductivity follows the equation proposed by ~\citetSI{Every2000}: 
\begin{gather}
    \sigma(T, c) = \sum n_i q_i \mu_i
\end{gather}
where $\sigma$ is ionic conductivity, $n_i$ is the number of ions, $q_i$ is the charge of the ion species $i$ and $\mu_i$ is the mobility. In relatively medium and high salt concentration, the equation can be extended by defining an effective number of free ions. Based on ~\citetSI{Zhang2020}'s work, the number of free ions $n_i$ is a function of the electrolyte concentration $c$. As the concentration increases, there is an increasing number of ion associations. Consequently, the number of free ions will not increase linearly as the concentration arises. Their work assumed the following relation between the number of free ions and salt concentration~\citeSI{Zhang2020} and we use it in this work:
\begin{gather}
    n_i = Ac^{n_1}
\end{gather}
In terms of mobility $\mu$, it is empirically described by an exponential relationship that depends on both temperature and concentration. For the temperature dependence, the VFT equation is widely used~\citeSI{ngai2011relaxation} and has been validated by extensive experimental studies~\citeSI{Ding2001, Ding2018, Hirotsugu2015}. As a result, the mobility observes the following equation:
\begin{gather}
    \mu = \mu_0e^{-\frac{E_a}{T-T_0}}
\end{gather}
where $\mu_0$ is the mobility when temperature is infinite. $T_0$ is physically correlated to the glass transition temperature of the electrolytes and $E_a$ is the activation energy for electrical conduction. In~\citetSI{Fu2018}'s work, the activation energy correlates to the concentration with the below equation based on quasi-lattice theory:
\begin{gather}
    E_a = (ac^{-0.5} +b)c+d
\label{si_eq: activation_energy_orig}
\end{gather}
where $a$, $b$ and $d$ are system-dependent parameters. The exponent -0.5 is chosen based on experimental observations of the [BMIM][TFSI]-PC/GBL mixture: as concentration increases, the activation energy increases, but the slope of the curve decreases. However, we found that in other systems like LiClO$_4$ in $\gamma$-butyrolactone~\citeSI{Chagnes2001}, the slope in activation energy-concentration plot can also increases. To accommodate a more flexible dependence of $E_a$ on concentration, we adopt the following modified expression:
\begin{gather}
    E_a = Bc^{n_2} + D
\end{gather}
where $B$ and $D$ are adjustable parameters, and $n_2$ is constrained to be greater than zero only. 

In terms of viscosity, we incorporate it following Walden’s rule to provide prior intuition for predicting conductivity as the conductivity generally decreases as viscosity increases. If an electrolyte molecule is unseen by the model, its viscosity can serve as a prior to infer its potential impact on the overall electrolyte conductivity. To calculate the viscosity of mixed solvents in the electrolyte, we assume the mixture behaves as an ideal solution. Although this assumption may introduce inaccuracies, the empirical equation incorporates learnable parameters ($A$) that can correct any resulting deviations. As a result, incorporating such a term in the equation should not negatively impact the prediction. The viscosity of the mixture is then determined based on the viscosities of individual solvent molecules, which are predicted during molecular pretraining. The viscosity is described with Arrhenius blending rule~\citeSI{GRUNBERG1949, Boehm2022}:
\begin{gather}
    ln(\eta) = \sum r_iln(\eta_i)
\end{gather}
where $r_i$ is the molar ratio of each solvent molecule and $\eta_i$ is the predicted viscosity of each single molecule. By plugging in all components, the final ionic conductivity is defined with the following equation:
\begin{gather}
     \sigma(T, c) = \frac{A}{\eta}c^{n_1}e^{-\frac{B \times c^{n_2} + D}{T - T_0}} 
\label{si_eq:emp_relation}
\end{gather}
where $\sigma$ is the conductivity, $c$ is salt concentration, $T$ is the temperature, $\eta$ is the viscosity, $A$, $B$, $D$, $n_1$, $n_2$ and $T_0$ are all learnable parameters, dependent on the electrolyte embedding ($\bm{h}_e$). During the training of the model, we actually use the logarithmic form of Eq.~\ref{si_eq:emp_relation} for better training stability given that exponential term can easily explode or vanish:
\begin{gather}
    \text{ln}(\sigma(T, c)) = \text{ln}(A) - \text{ln}(\eta) + n_1 \text{ln}(c) - \frac{B \times c^{n_2} + D}{T - T_0} \\
     \text{ln}(A) = \text{MLP}(\bm{h}_e,  \text{act=softplus}) \\
     B = \text{MLP}(\bm{h}_e,  \text{act=softplus}) \\
     D = \text{MLP}(\bm{h}_e, \text{act=softplus}) \\
     n_1 = \text{MLP}(\bm{h}_e, \text{act=softplus}) \\
     n_2 = \text{MLP}(\bm{h}_e,  \text{act=softplus}) \\
     T_0 = \text{MLP}(\bm{h}_e, \text{act=sigmoid}) 
\label{si_eq:empirical_equation_log}
\end{gather}
where the activation functions of last layer of each MLP readout block are listed above, which constrains the range of learnable parameters within the empirical relation. The reason we use \enquote{sigmoid} function for $T_0$ is that we only consider cases when the working temperature $T$ is higher than $T_0$ or glass transition temperature in this work. 

\begin{figure}[htbp!]
\centering
\includegraphics[width=\linewidth]{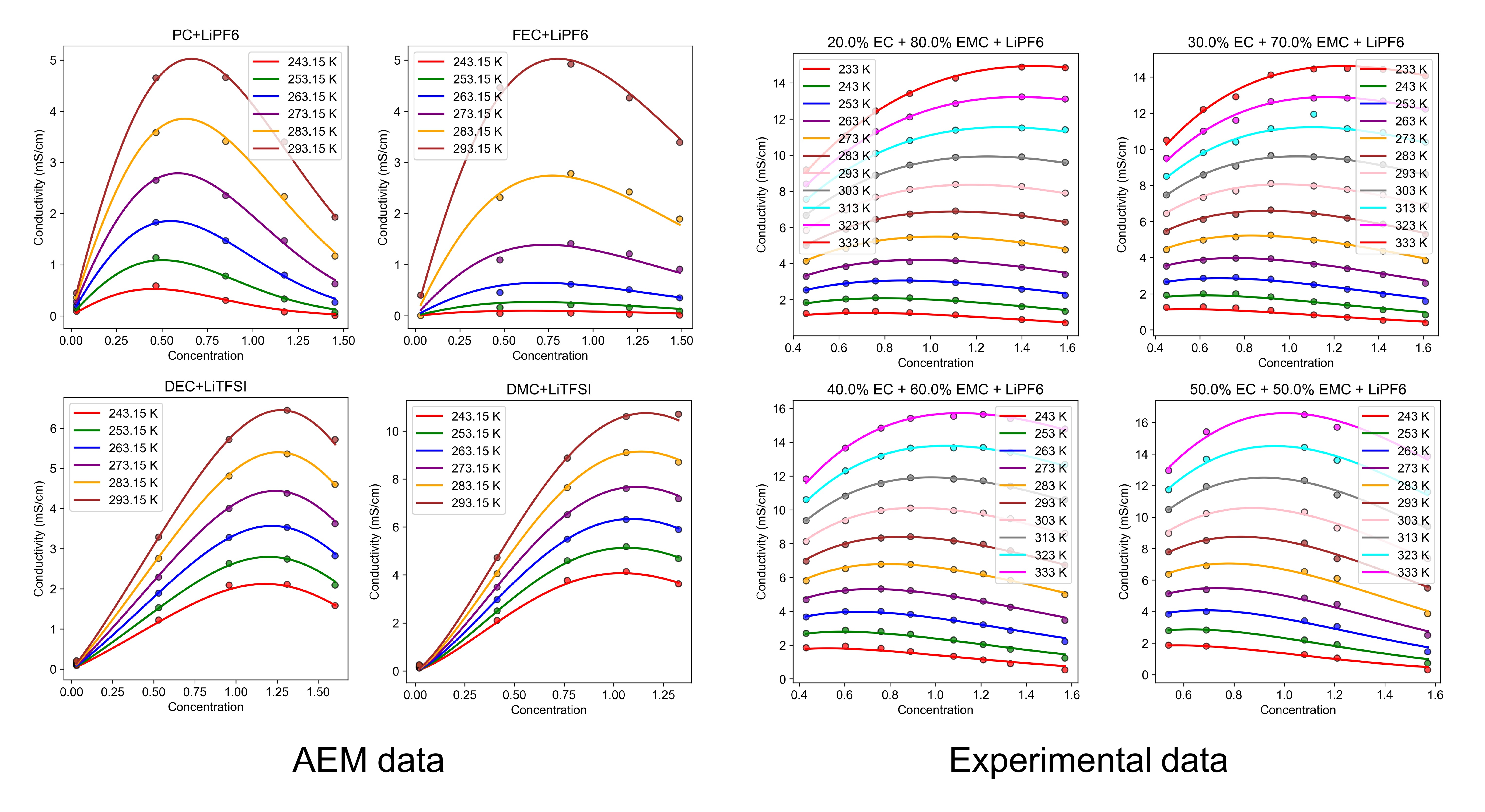}
\caption{\textbf{Empirical equation validation.} AEM and experimental data of various electrolyte formulation systems are utilized to validate our chosen empirical relation. } 
\label{si_fig:emp_relation}
\end{figure}

The empirical equation is further validated using AEM~\citeSI{Zhu2024} and experimental data~\citeSI{Ding2001}. Given six fitting parameters (except viscosity) in the empirical relation, we select only electrolyte systems that contain more than 30 data points spanning different temperatures and concentrations for validation of our propose empirical relation. To demonstrate, we select AEM data from four different binary mixtures and test experimental data for EC/EMC/LiPF$_6$ electrolyte mixture with varying component ratios (Fig.~\ref{si_fig:emp_relation}). For curve fitting, we utilized the \enquote{curve\_fit} function in SciPy with a least-squares method~\citeSI{2020SciPy-NMeth}. All the learnable parameters are constrained to be greater than 0 and the maximum number of function evaluations during curve fitting is set to 10000. 

An ablation study was further conducted to illustrate the necessity of parameters in our empirical equation. Given the mathematical form of the empirical equation, we removed $n_1$, $n_2$, and $D$ respectively to evaluate their impact on the fitting accuracy. As Table~\ref{si_tab:empirical_relation} reveals, the error (MSE) increases considerably when any of these parameters are removed. These results demonstrate that each parameter is critical for accurately capturing the empirical relationship. While several empirical relations can describe how concentration and temperature influence conductivity, our equation performs well, achieving accurate predictions across more than 10,000 experimental measurements.

\section{Generative electrolyte design}
\subsection{Synthetic dataset}
We use a synthetic dataset generated by our predictive model to train conditional diffusion model. The synthetic dataset provides more balanced coverage of the electrolyte design space, with a particular emphasis on multi‑component systems that are really used in commercial battery systems. The dataset contains electrolyte formulations with various numbers of solvents listed as follows: (1) 1 solvent + 2 salts (62 data including all possible solvents); (2) 2 solvents + 2 salts (1891 data including all possible binary combinations of solvents); (3) 3 solvents + 2 salts (2805 data which is 10\% of the rest of data); (4) 4 solvents + 2 salts (5609 data which is 20\% of the rest of data); (5) 5 solvents + 2 salts (8414 data which is 30\% of the rest of data); (6) 6 solvents + 2 salts (11219 data which is 40\% of the rest of data). We selected six solvents and two salts as the maximum to maintain consistency with the computational dataset, and eight components are generally sufficient to formulate a commercially viable electrolyte system. The molar ratios for solvents and salts are randomly sampled between 0 and 1 and normalized within solvents and salts, respectively. 
The overall salt concentration is fixed at 0.1 in terms of molar ratio, which is typically near the ideal concentration corresponding to peak conductivity, and the temperature is set to 25 \textdegree{}C as the room temperature. 
We include all possible unary and binary combinations of solvent molecules in the dataset and randomly sample multi-component systems based on their combinatorial count to ensure comprehensive coverage of the formulation space.

\subsection{Conditional diffusion model}
We utilize a diffusion model to generate electrolyte formulations given specific target properties. Electrolyte embeddings ($\bm{h}_e$) serve as the target output, reducing generation costs given their low dimensionality while ensuring permutation invariance. More specifically, the conditional generation is achieved based on the DDPM model~\citeSI{ho2020denoisingdiffusionprobabilisticmodels}. In a DDPM, the forward process (diffusion process) defines a Markov chain that gradually adds Gaussian noise to the input data according to a variance schedule $\beta_1$, $\beta_2$, ..., $\beta_T$:
\begin{gather}
    q(\bm{x}_{1:T}|x_0) = \prod_{t=1}^T q(\bm{x}_t|\bm{x}_{t-1}) \\
    q(\bm{x}_t|\bm{x}_{t-1}) = \mathcal{N}(\bm{x}_t; \sqrt{1-\beta_t}\bm{x}_{t-1},\beta_t\bm{I})
\end{gather}
where $\bm{x}_t$ is the data at time $t$, $\bm{I}$ is identity matrix/tensor. As time $t$ increases, the data become more and more noisy. The key objective is to reverse this diffusion process, enabling the generation of data from random noise. The reverse process is defined as: 
\begin{gather}
    p_\theta(\bm{x}_{t-1} | \bm{x}_t) = \mathcal{N}(\bm{x}_{t-1}; \bm{\mu}_\theta(\bm{x}_t, t), \bm{\Sigma}_\theta(\bm{x}_t, t)) \label{si_eq: reverse_prob} \\
    \bm{\mu}_\theta(\bm{x}_t, t) = \frac{1}{\sqrt{\alpha_t}} \left(\bm{x}_t - \frac{\beta_t}{\sqrt{1 - \bar{\alpha}_t}} \bm{\epsilon}_{\theta}(\bm{x}_t, t) \right) \\
    \Sigma_\theta(\bm{x}_t, t) = \frac{1-\bar{\alpha}_{t-1}}{1-\bar{\alpha}_t} \beta_t\bm{I}
\end{gather}
where $\alpha_t = 1 - \beta_t$, $\bar{\alpha}_t = \prod_{i=1}^t \alpha_i$, $\bm{\epsilon}_{\theta}$ is a functional approximator to predict added noise $\bm{\epsilon}$ during the diffusion process from $\bm{x}_t$. The functional approximator is a 1D U-Net model~\citeSI{unet2015} we adopted from an open-source Github repository~\citeSI{denoisingdiffusion2023}. The 1D U-Net contains a downsampling encoder, a middle bottleneck block and an upsampling decoder with self-attention mechanism. We used a base hidden dimension of 64, and for each upsampling or downsampling layer, the hidden dimension is doubled or halved relative to the previous layer. The number of attention head is set to 4 and the dimension is 16 for self-attention blocks in the U-Net. 

To achieve conditional generation, the input conditions are first processed through an MLP block to produce a high-dimensional conditional embedding vector. This embedding is then treated similarly as the time embedding in a diffusion model, where it modulates the input via Feature-wise Linear Modulation (FiLM)~\citeSI{Perez2017}. In the process, the model dynamically scales and shifts feature activations based on the time and conditioning variables, enabling effective control over the generated output:
\begin{gather}
    \bm{x} = (1+\gamma(v, t))\bm{x}+\beta(v, t) \\
    \gamma(v, t), \beta(v,t) = \text{MLP}(v) + \text{MLP}(t)
\end{gather}
where $v$ is the conditional variable, including conductivity and anion ratio in this work. Both the time and conditional embedding dimensions are set to 256. 

In terms of other hyperparameters, the number of diffusion step is 1000. A cosine scheduler is implements for $\beta_t$. Training is performed with a batch size of 256, using a 70/30 train-test split to minimize the MSE loss of predicted noises. Early stopping with a patience of 20 epochs is applied to optimize the training process.

\subsection{Electrolyte decoder}\label{si_sec:electrolyte_decoder}
A decoder is developed to convert generated electrolyte embeddings back into their original electrolyte formulations. We first transfer the electrolyte embedding to a set of molecular embeddings of solvents and salts, and then match each molecular embedding to our chemical vocabulary to identify the corresponding molecules. The molar ratios of each molecule are also predicted from the electrolyte embedding. Based on the matching results and predicted molar ratios, we derive the \enquote{BoM} vector for each electrolyte formulation. There are two main reasons for retaining molecular embeddings during the decoding process. First, this approach can be extended for novel electrolyte molecule discovery by incorporating an additional molecular decoder capable of converting molecular embeddings back into molecular SMILES. Second, molecular embeddings possess valuable chemical information, enhancing the overall expressivity of the workflow.  

To generate high-dimensional molecular embeddings for solvents and salts from the low-dimensional electrolyte embedding, we utilize a learnable seed vectors $\bm{S} \in \mathbb{R}^{N\times d}$ ($N$ is the number of solvents and salts, $d$ is the dimension of molecular embedding)~\citeSI{Lee2019} to extract molecular embedding at each position given the electrolyte embedding as the context. The steps are as follows:  
\begin{gather}
    \bm{c} = \text{MLP}(\bm{h}_e) \in \mathbb{R}^{1 \times d} \\
    \bm{C} = [\bm{c}, \bm{c}, \bm{c}, ...] \in \mathbb{R}^{N \times d} \\
    \bm{S}^{(0)} = \bm{S} \\
    Q^{(L)} = W_Q^{(L)} \bm{S}^{(L)} + B_Q^{(L)} \\
    K^{(L)} = W_K^{(L)} \bm{C} + B_K^{(L)} \\
    V^{(L)} = W_V^{(L)} \bm{C} + B_V^{(L)} \\
    \bm{S}^{(L+1)} = \text{softmax}(\frac{Q^{(L)}(K^{(L)})^T}{\sqrt{d_{K^{(L)}}}}) V^{(L)} \in \mathbb{R}^{N \times d}  \\
    \bm{S}^{(F)} = \bm{h}_m \\
    \bm{r} = \text{softmax}(\text{MLP}(\text{concat(}\bm{h}_m, \bm{C})))
\end{gather}
where $\bm{C}$ is the context embedding based on the electrolyte embedding, $\bm{S}^{(L)}$ denotes the output at layer $L$,  $\bm{S}^{(0)}$ is the input learnable query and $\bm{S}^{(F)}$ is the output of last layer which corresponds to the molecular embeddings. The solvents and salts are processed separately using the above procedure. 

With the molecular embeddings, we further match each embedding to molecules within our chemical vocabulary. This matching is performed by computing the L2 distance between embedding vectors. Since a hard indexing operation would disrupt the computational graph and hinder back propagation, we employ a soft matching mechanism instead:
\begin{gather}
    \bm{p}: p_{i, j} =\text{softmax}(-\lambda d_{i, j})
\end{gather}
where $p_{i, j}$ is the probability of \enquote{Mol i} is same/similar as \enquote{Mol j}, $d_{i, j}$ is the L2 distance between embedding vectors of these two molecules, $\lambda$ is a parameter to control the sharpness of the matching mechanism. In other words, a larger $\lambda$ makes the matching process more selective, increasing the likelihood of collapsing onto a single molecule. In this work, we set $\lambda$ to 1000 to make the matching process closer to a hard indexing operation while avoiding numerical instability that may arise when $\lambda$ is too large. 

The final BoM vector is then obtained based on the inner product of matching probability and predicted molar ratios:
\begin{gather}
    \bm{v}_{\text{BoM}} = \bm{p}^T \cdot \bm{r}
\end{gather}
The training of the decoder is performed by minimizing the MSE loss of predicted and actual $\bm{v}_{\text{BoM}}$.

\subsection{Evaluation metrics}
The three evaluation metrics we utilized to examine the performances of generation are MAPE, formulation diversity and molecular diversity. The definitions of these metrics are listed as follows:
\begin{gather}
    \text{MAPE} = \sum (|\frac{y_{c}^{\text{pred}} - y_{c}^{\text{target}}}{y_{c}^{\text{target}}}| + |\frac{y_{a}^{\text{pred}} - y_{a}^{\text{target}}}{y_{a}^{\text{target}}}|)
\end{gather}
where $y_{c}^{\text{pred}}$ and $y_{c}^{\text{target}}$ are predicted and target conductivities respectively, $y_{a}^{\text{pred}}$ and $y_{a}^{\text{target}}$ are predicted and target anion ratios. 

The formulation diversity is defined based on the average minimal pairwise L1 distance between $\bm{v}_{\text{BoM}}$ of generated formulations, showing the difference between generated formulations. In other words, it is the average distance from each generated formulation to its nearest neighbor:
\begin{gather}
    \text{formulation diversity} = \frac{1}{N} \sum_{i=1}^{N} \min_{\substack{j = 1 \\ j \ne i}}^{N} \sum_{k=1}^{d} |\bm{v}_{BoM, i}^{(k)} - \bm{v}_{BoM, j}^{(k)}|
\end{gather}
where $d$ is the dimension of $\bm{v}_{\text{BoM}}$. $N$ is the number of generated samples. We use the L1 distance to evaluate differences between electrolyte formulations, as it not only captures variations in the molar ratios of shared molecular species but also imposes a large penalty when there is no overlap in molecular components. For example, if the molar ratio of EC increases from 0.2 to 0.4 in a formulation, the L1 distance between the original and modified formulations is 0.2. In contrast, if 20\% EC is replaced with 20\% DMC, the L1 distance increases to 0.4. In addition, the L1 distance is inherently normalized, since the molar ratios sum to one.

The molecular diversity is defined based on the Shannon entropy/relative entropy of occurrence frequency of different component molecules calculated using Scipy~\citeSI{2020SciPy-NMeth}: 
\begin{gather}
    \text{molecular diversity} = -\sum_{i=1}^{N} p_i \log(p_i)
\end{gather}
where $p_i$ is the occurrence frequency of component molecule $i$. The frequency is normalized during the calculation. This metric captures the diversity of electrolyte formulations considering the molecular species. For instance, in high-conductivity electrolyte systems, acetonitrile (AN) is widely seen which can be reflected by this score. 
\subsection{Classifier-guided diffusion}
To generate electrolyte formulations that satisfy base formulation constraints, we employ a CGD-based method. The CGD method explicitly calculates gradients during the sampling process (reverse process) given predicted label $y$ from the classifier with respect to the noisy data. Instead of sampling from the Gaussian distribution at each time step $t$ as Eq.~\ref{si_eq: reverse_prob} shows, it shifts the mean value $\bm{\mu}_\theta(\bm{x}_t, t)$ towards the target class using the gradients:
\begin{gather}
    \bm{\mu}_\theta(\bm{x}_t, y, t) = \bm{\mu}_\theta(\bm{x}_t, t) + s\sum   \nabla_{\bm{x}_t} \log p_{\phi}(y | \bm{x}_t) 
\end{gather}
where $s$ is a gradient scale that adjusts the strength of classifier-guided generation. $\phi$ is the classifier function. A larger $s$ results in more sharp guidance towards the target constraint. 

To accommodate different base formulation scenarios, we formulate the classifier function as follows:
\begin{gather}
    \bm{v}_{\text{BoM}} = \text{Decoder}(\bm{x}_t) \\
    \phi(\bm{x}_t, \bm{b}, \bm{\tau}, \bm{m}) =\sum_{i=1}^{n_m} -\text{ReLU}((\bm{b}[i]-\bm{v}_{\text{BoM}}[\bm{m}_i]) \times \bm{\tau}_i) 
\end{gather}
where $\bm{m}$ is a list of molecule indices which require control on molar ratios, $\bm{b}$ is the corresponding bound values of molar ratio, $\bm{\tau}$ indicates (\enquote{-} or \enquote{+}) whether the $\bm{\tau}$ stores the upper or lower bounds. $n_m$ is the number of molecules that require controls within the base formulation. Below is an example of representing a base formulation with $\bm{m}$, $\bm{b}$ and $\bm{\tau}$:
\begin{gather}
    \bm{m} = [0, 2],  \bm{b} = [0.2, 0.4],  \bm{\tau} = [1, -1]  
    \Leftrightarrow \text{\enquote{Mol 0}} > 20\% \text{ and \enquote{Mol 2}} < 40\%
\end{gather}
More specifically, in the above example, $\bm{m}$ stores the indices of two molecules in the base formulation, namely \enquote{Mol 0} and \enquote{Mol 2},  which are referenced based on our molecular dictionary. The vector $\bm{b}$ provides the molar ratio thresholds for these molecules, with 0.2 for \enquote{Mol 0} and 0.4 for \enquote{Mol 2}. Finally, in $\bm{\tau}$, a value of \enquote{1} indicates that the molar ratio of the corresponding molecule should be greater than its threshold (lower bound), while \enquote{-1} signifies that the molar ratio should be less than its threshold (upper bound). 

This classifier function allows us to handle various base formulation requirements while also enabling precise control over the molar ratios of component molecules. Specifically, we can achieve this by setting both upper and lower bounds for a given molecule. Since small variations in molar ratios typically have minimal impact on key electrolyte properties, such as conductivity, we can apply a relatively soft constraint.

Finally, we utilize a negative ReLU function to ensure proper gradient behavior during sampling. Above all, the gradient should be zero when the condition is satisfied. In other words, one side of the activation function should be constant as in ReLU function. Second, when the condition is not met, the gradient should be positive; thus, a negative version of ReLU is used. Third, since the ReLU function maintains a constant gradient when the condition is unsatisfied, we do not need to adjust the scale of molar ratios to avoid gradient vanishing or exploding issues. The gradient strength can be easily adjusted via the gradient scale, offering flexibility in guidance control. The gradient scales $\bm{s}$ are chosen between 0, 1, 10, 100, 1000 and 10000. When gradient scale is 0, the classifier does not impose any constraint to the generation which falls back to conditional generation with only conductivity and anion ratio targets. We found that for both base formulation constraints we tested in this study --- \enquote{EC $>$ 20\%} and \enquote{EC $>$ 20\%, DMC $>$ 20\% and EMC $>$ 20\%}, the success rates are highest when $\bm{s} = 1000$. 

For the decoder used in the CGD method, we retrain it with noisy electrolyte embeddings to predict the corresponding BoM vectors. The noisy data are sampled directly from the forward process in the diffusion model during training. As expected, the decoder trained on noisy data exhibits a higher loss compared to the model trained on the final electrolyte embedding (Section~\ref{si_sec:electrolyte_decoder}). However, this training strategy enhances the model’s ability to establish a more robust linkage between noisy data during sampling and the final electrolyte formulation which is represented by $\bm{v}_{\text{BoM}}$.

\section{Experimental validation}
The 18 formulations we used for evaluation are selected out of 29 formulations measured by experiments, including 8 reference single-solvent formulation and 21 generated multi-solvent formulations. 
During the experiments, we observed that the reference DOL/LiFSI binary electrolyte and 10 generated multi-component electrolytes either did not exhibit distinct peaks in their Raman spectra or displayed unusually high viscosity. These data were considered unreliable and excluded from further evaluation. 
As for the potential causes of these unreliable data, the DOL/LiFSI binary solution is known to undergo spontaneous polymerization reactions~\citeSI{jie2022molecular}. The absence of distinct Raman features or ultrahigh viscosity in the 10 generated electrolytes may similarly result from polymerization, side reactions, or salt crystallization. 

Chemical stability, reactivity, and solubility of electrolytes are not currently accounted for in the model’s predictions and generative processes, and they are inherently complicated. For example, the polymerization of DOL can be suppressed in the presence of certain chemical species, allowing the solution to remain stable~\citeSI{jie2022molecular}. Furthermore, relevant studies and experimental data remain limited. As a potential path forward, future work could incorporate base formulation constraints into our generative workflow, like we did in this work, to embed domain knowledge and better address issues related.
\clearpage

\begin{table}[htbp!]
\centering

\resizebox{\columnwidth}{!}{%
\begin{tabular}{@{}ccc@{}}
\toprule
\textbf{Notation} & \textbf{Data representation} & \textbf{Shape} \\ \midrule
$\bm{X}$        & padded molecular embeddings of solvents and salts within the electrolyte.        & (24, 64)     \\
$\bm{x}/\bm{x}_t$        & noisy data of electrolyte embedding during diffusion and denoising process.        & (384, )     \\ 
$\bm{h}_m$   & molecular embedding of one single molecule. & (64, ) \\
$\bm{h}_e$   & electrolyte embedding of one formulation. & (384, ) \\
$\bm{v}_{\text{BoM}}$ & \enquote{Bag-of-Molecules} vector to represent one electrolyte formulation. & (vocab\_size, ) \\ \bottomrule
\end{tabular}%
}
\caption{\textbf{Data representation notation}. The table presents the key data representations for both molecules and electrolyte formulations, along with their corresponding notations and common dimensions used in this work. \enquote{vocab\_size} refers to the number of available solvent and salt molecules within our chemical database.}
\label{si_tab:data_repr}
\end{table}

\begin{table}[htbp!]
\centering
\resizebox{\columnwidth}{!}{%
\begin{tabular}{@{}ccccccccc@{}}
\toprule
\textbf{Data split} &
  \multicolumn{2}{c}{no F/with F} &
  \multicolumn{2}{c}{no P/with P} &
  \multicolumn{2}{c}{no S/with S} &
  \multicolumn{2}{c}{high viscosity/low viscosity} \\ \midrule
\multicolumn{1}{l}{\textbf{Method}} &
  \begin{tabular}[c]{@{}c@{}}Computational \\ pretraining\end{tabular} &
  \begin{tabular}[c]{@{}c@{}}No \\ pretraining\end{tabular} &
  \begin{tabular}[c]{@{}c@{}}Computational \\ pretraining\end{tabular} &
  \begin{tabular}[c]{@{}c@{}}No \\ pretraining\end{tabular} &
  \begin{tabular}[c]{@{}c@{}}Computational \\ pretraining\end{tabular} &
  \begin{tabular}[c]{@{}c@{}}No \\ pretraining\end{tabular} &
  \begin{tabular}[c]{@{}c@{}}With \\ viscosity\end{tabular} &
  \begin{tabular}[c]{@{}c@{}}No \\ viscosity\end{tabular} \\ \midrule
\bm{$R^2$} &
  \textbf{0.797} &
  0.491 &
  \textbf{0.901} &
  0.893 &
  \textbf{0.611} &
  0.238 &
  \textbf{0.559} &
  0.425 \\ \bottomrule
\end{tabular}%
}
\caption{\textbf{Generalization tests.} We evaluated how computational pretraining and the inclusion of viscosity in the empirical relation influence the generalization ability of our model. The label \enquote{no F / with F} indicate that the training set contains electrolytes with non-fluorinated solvents, while the test set contains electrolytes with fluorinated solvents. Similar notations are used for \enquote{no S/with S} and \enquote{no P/with P}. The label \enquote{high viscosity / low viscosity} corresponds to a scenario in which low-viscosity solvents including ethyl acetate (EA), methyl acetate (MA), and AN are excluded from the training set and included only in the test set. The $R^2$ score is computed after performing a linear fit between the predicted and actual conductivity values. }
\label{si_tab:generalization}
\end{table}

\begin{table}[htbp!]
\centering
\resizebox{\columnwidth}{!}{%
\begin{tabular}{@{}ccccccc@{}}
\toprule
\textbf{Property}  & $\bm{T_m}$                 & $\bm{T_b}$                   & $\bm{n_D}$                   & $\bm{n_D^{\text{liquid}}}$          & $\bm{pK_a}$                & $\bm{pK_b}$                 \\ \midrule
\textbf{Resource} &
  \begin{tabular}[c]{@{}c@{}}~\citetSI{Tetko2016}\\ ~\citetSI{Bradley2014}\\ CAS Covid dataset~\citeSI{cas_convid}\\ CRC handbook~\citeSI{crchandbook2004}\\ OPERA dataset~\citeSI{Mansouri2018}\end{tabular} &
  \begin{tabular}[c]{@{}c@{}}CRC handbook~\citeSI{crchandbook2004}\\ OPERA dataset~\citeSI{Mansouri2018}\end{tabular} &
  CRC handbook~\citeSI{crchandbook2004} &
  CRC handbook~\citeSI{crchandbook2004} &
  OPERA dataset~\citeSI{Mansouri2019} &
  OPERA dataset~\citeSI{Mansouri2019} \\ \midrule
\textbf{Property}  & $\boldsymbol{\varepsilon}$                  & $\bm{\gamma_s}$                   & $\bm{\rho}$              & $\bm{\eta}$                  & $\bm{P_{\text{vap}}}$            \\ \midrule
\textbf{Resource} &
  ~\citetSI{Bouteloup2019} &
  ~\citetSI{Wohlfarth1997} &
  CRC handbook~\citeSI{crchandbook2004} &
  \begin{tabular}[c]{@{}c@{}}~\citetSI{Goussard2020}\\ ~\citetSI{Chew2024}\end{tabular} &
  \begin{tabular}[c]{@{}c@{}}OPERA dataset~\citeSI{Mansouri2018}\\ ~\citetSI{Gharagheizi2012}\end{tabular} &
   \\ \bottomrule
\end{tabular}%
}
\caption{\textbf{Single-molecule dataset}. The table lists the data resources of different molecular properties.}
\label{si_tab:mono_dataset}
\footnotetext{$\bm{T_m}$: melting point; $\bm{T_b}$: boiling point; $\bm{n_D}$/$\bm{n_D^{\text{liquid}}}$: (liquid) refractive index; $\bm{\varepsilon}$: dielectric constant; $\bm{\gamma_s}$: surface tension, $\bm{\eta}$: viscosity; $\bm{P_{\text{vap}}}$: vapor pressure.}
\footnotetext{\textbf{Total number of data}: $\bm{T_m}$: 184921; $\bm{T_b}$: 7464; $\bm{n_D}$: 4252; $\bm{n_D^{\text{liquid}}}$: 2592; $\bm{pK_a}$: 2568; $\bm{pK_b}$: 3283, $\bm{\varepsilon}$: 1286; $\bm{\gamma_s}$: 14317; $\bm{\rho}$: 5181; $\bm{\eta}$: 3405; $\bm{P_{\text{vap}}}$: 32184.}
\end{table}

\begin{table}[htbp!]
\centering
\resizebox{\columnwidth}{!}{
\begin{tabular}{@{}ccccccc@{}}
\toprule
\multirow{2}{*}{\textbf{Model}}  & $\bm{T_m}$                 & $\bm{T_b}$                   & $\bm{n_D}$                   & $\bm{n_D^{\text{liquid}}}$          & $\bm{pK_a}$                & $\bm{pK_b}$                \\ \cmidrule(l){2-7} 
                        & $R^2$/\text{MAE}               & $R^2$/\text{MAE}               & $R^2$/\text{MAE}               & $R^2$/\text{MAE}              & $R^2$/\text{MAE}               & $R^2$/\text{MAE}              \\ \midrule
Morgan fingerprint + NN & 0.506/0.045          & 0.625/0.035          & 0.768/0.026          & 0.789/0.029          & 0.537/0.055          & 0.541/0.036          \\
Atom/bond feature + GAT & 0.629/0.039          & 0.799/0.020          & 0.825/0.014          & 0.892/0.014          & 0.651/0.044          & 0.750/0.027          \\
Atom/bond feature + EGT & \textbf{0.649/0.038} & \textbf{0.888/0.015} & \textbf{0.878/0.012} & \textbf{0.932/0.013} & \textbf{0.819/0.033} & \textbf{0.852/0.023} \\ \midrule
\multirow{2}{*}{\textbf{Model}}  & $\boldsymbol{\varepsilon}$                  & $\bm{\gamma_s}$                   & $\bm{\rho}$              & $\bm{\eta}$                  & $\bm{P_{\text{vap}}}$                 &                      \\ \cmidrule(l){2-7} 
                        & $R^2$/\text{MAE}               & $R^2$/\text{MAE}               & $R^2$/\text{MAE}               & $R^2$/\text{MAE}               & $R^2$/\text{MAE}               &                      \\ \midrule
Morgan fingerprint + NN & 0.744/0.062          & 0.600/0.048          & 0.747/0.027          & 0.530/0.098          & 0.558/0.077          &                      \\
Atom/bond feature + GAT & 0.982/0.019          & 0.900/0.019 & 0.842/0.014 & 0.932/0.030          & 0.927/0.022          &                      \\
Atom/bond feature + EGT & \textbf{0.991/0.014} & \textbf{0.958/0.013}          & \textbf{0.880/0.013}          & \textbf{0.949/0.027} & \textbf{0.967/0.015} &                      \\ \bottomrule
\end{tabular}
}

\caption{\textbf{Model performance comparison for molecular property predictions.} The best performance of each property among the three models is highlighted in bold. All properties are normalized based on the minimum and maximum values within the dataset. For \enquote{$\epsilon$}, \enquote{$\eta$} and \enquote{$P_{\text{vap}}$}, the properties are log-normalized due to their long-tail distribution or exponential dependence on temperature.}
\label{si_tab:pretrain_comparison}
\end{table}

\begin{table}[htbp!]
\centering
\resizebox{\columnwidth}{!}{%
\begin{tabular}{@{}ccccc@{}}
\toprule
Empirical form    & $Ac^{n_1}e^{-\frac{B \times c^{n_2} + D}{T - T_0}}$   & $Ace^{-\frac{B \times c^{n_2} + D}{T - T_0}}$   & $Ac^{n_1}e^{-\frac{B \times c + D}{T - T_0}}$   & $Ac^{n_1}e^{-\frac{B \times c^{n_2}}{T - T_0}}$  \\ 
Parameters & $(A, B, D, n_1, n_2, T_0)$ & $(A, B, D, n_2, T_0)$ & $(A, B, D, n_1, T_0)$ & $(A, B, n_1, n_2, T_0)$ \\ \midrule
AEM data~\citeSI{Zhu2024}          & \textbf{0.00504} & 0.01283 & 0.01237 & 0.02458 \\
Experimental data~\citeSI{Ding2001} & \textbf{0.01326} & 0.01663 & 0.02108 & 0.09039 \\ \bottomrule
\end{tabular}%
}
\caption{\textbf{Ablation study of empirical relations.} We compare our empirical equations with modified versions where one parameter is removed using both AEM and experimental data. The values represent the mean MSEs ((mS/cm)$^2$) across different electrolyte formulations of each empirical relation after parameter fitting. Our selected equation exhibits better fitting accuracy compared to these modified versions. }
\label{si_tab:empirical_relation}
\end{table}

\begin{table}[htbp!]
\centering
\resizebox{\columnwidth}{!}{%
\begin{tabular}{@{}ccccc@{}}
\toprule
\textbf{Molecule index} & \textbf{Molecule abbreviation} & \textbf{Molecule full name} & \textbf{Molecule SMILES} \\ \midrule
1  & EC    & ethylene carbonate        & C1COC(=O)O1     \\
2  & PC    & propylene carbonate       & CC1COC(=O)O1    \\
3  & MA    & methyl acetate            & CC(=O)OC        \\
4  & EA    & ethyl acetate             & CCOC(=O)C       \\
5  & DEC   & diethyl carbonate         & CCOC(=O)OCC     \\
6  & DOL   & 1,3-dioxolane             & C1COCO1         \\
7  & THF   & tetrahydrofuran           & C1CCOC1         \\
8  & METHF & 2-Methyltetrahydrofuran   & CC1CCCO1        \\
9  & DMM   & dimethoxymethane          & COCOC           \\
10 & DEM   & diethoxymethane           & CCOCOCC         \\
11 & FEC   & fluoroethylene carbonate  & C1C(OC(=O)O1)F  \\
12 & MTFA  & methyl trifluoroacetate   & COC(=O)C(F)(F)F \\
13 & EMC   & ethyl methyl carbonate    & CCOC(=O)OC      \\
14 & EFA   & ethyl fluoroacetate       & CCOC(=O)CF      \\
15 & DME   & 1,2-dimethoxyethane       & COCCOC          \\
16 & EGDEE & 1,2-diethoxyethane        & CCOCCOCC        \\
17 & EP    & ethyl propionate          & CCC(=O)OCC      \\ \bottomrule
\end{tabular}%
}
\caption{\textbf{Solvent molecule list used in this work}. The list provides the molecular information of solvents used in formulations of this work.}
\label{si_tab:mol_name}
\end{table}

\begin{table}[htbp!]
\centering
\resizebox{\columnwidth}{!}{%
\begin{tabular}{@{}ccccc@{}}
\toprule
  \begin{tabular}[c]{@{}c@{}}\textbf{Formulation} \\ \textbf{index}\end{tabular} &
  \textbf{Components} &
  \textbf{Molar ratios} &
  \begin{tabular}[c]{@{}c@{}}\textbf{Conductivity} \\ \textbf{(mS/cm)}\end{tabular} &
  \begin{tabular}[c]{@{}c@{}}\textbf{Peak wavenumber}\\ \textbf{(cm$^{-1}$)}\end{tabular} \\ \midrule
1                 & EC/MA/EA/DOL/DEM/FEC/LiPF$_6$/LiFSI     & 9\%/4\%/18\%/13\%/41\%/6\%/2\%/7\%   & 18.77                & 713.1                 \\
2                 & PC/MA/EA/DOL/DEM/DMM/LiPF$_6$/LiFSI     & 3\%/21\%/22\%/10\%/15\%/19\%/4\%/6\% & 17.32                & 734.3                 \\
3  & MA/EA/DOL/MTFA/LiPF$_6$/LiFSI        & 22\%/17\%/42\%/9\%/1\%/9\%     & 16.75 & 713.1 \\
4  & MA/DEC/EA/EMC/LiFSI               & 19\%/5\%/43\%/7\%/10\%         & 15.95 & 723.2 \\
5  & THF/METHF/EMC/EFA/DME/LiPF$_6$/LiFSI & 43\%/2\%/3\%/5\%/38\%/1\%/8\%  & 15.92 & 711.1 \\
6  & MA/EA/EFA/MTFA/LiFSI              & 48\%/3\%/5\%/34\%/10\%         & 15.78 & 727.2 \\
7  & MA/EA/EGDEE/DEM/LiFSI             & 34\%/46\%/3\%/7\%/10\%         & 15.59 & 717.1 \\
8  & EA/DOL/DEM/EP/LiPF$_6$/LiFSI         & 7\%/32\%/25\%/26\%/2\%/8\%     & 14.66 & 712.1 \\
9 & EC/EA/THF/EGDEE/METHF/FEC/LiFSI   & 3\%/14\%/4\%/21\%/7\%/42\%/9\% & 12.79 & 724.2 \\ 
10 & MA/EA/DOL/DEM/LiFSI   & 11\%/36\%/27\%/16\%/10\% & 11.99 & 719.2 \\ \bottomrule
\end{tabular}%
}
\caption{\textbf{Generated formulations for experimental validation with target conductivity.} The table presents the component details and molar ratios of the tested formulations from our generative model. The peak wavenumbers are generally higher for formulations containing LiPF$_6$ as PF$_6^{-1}$ contributes a vibrational peak at a higher wavenumber compared to FSI$^{-1}$. }
\label{si_tab:exp_formulations}
\end{table}

\clearpage

\begin{figure}[htbp!]
\centering
\includegraphics[width=0.85\linewidth]{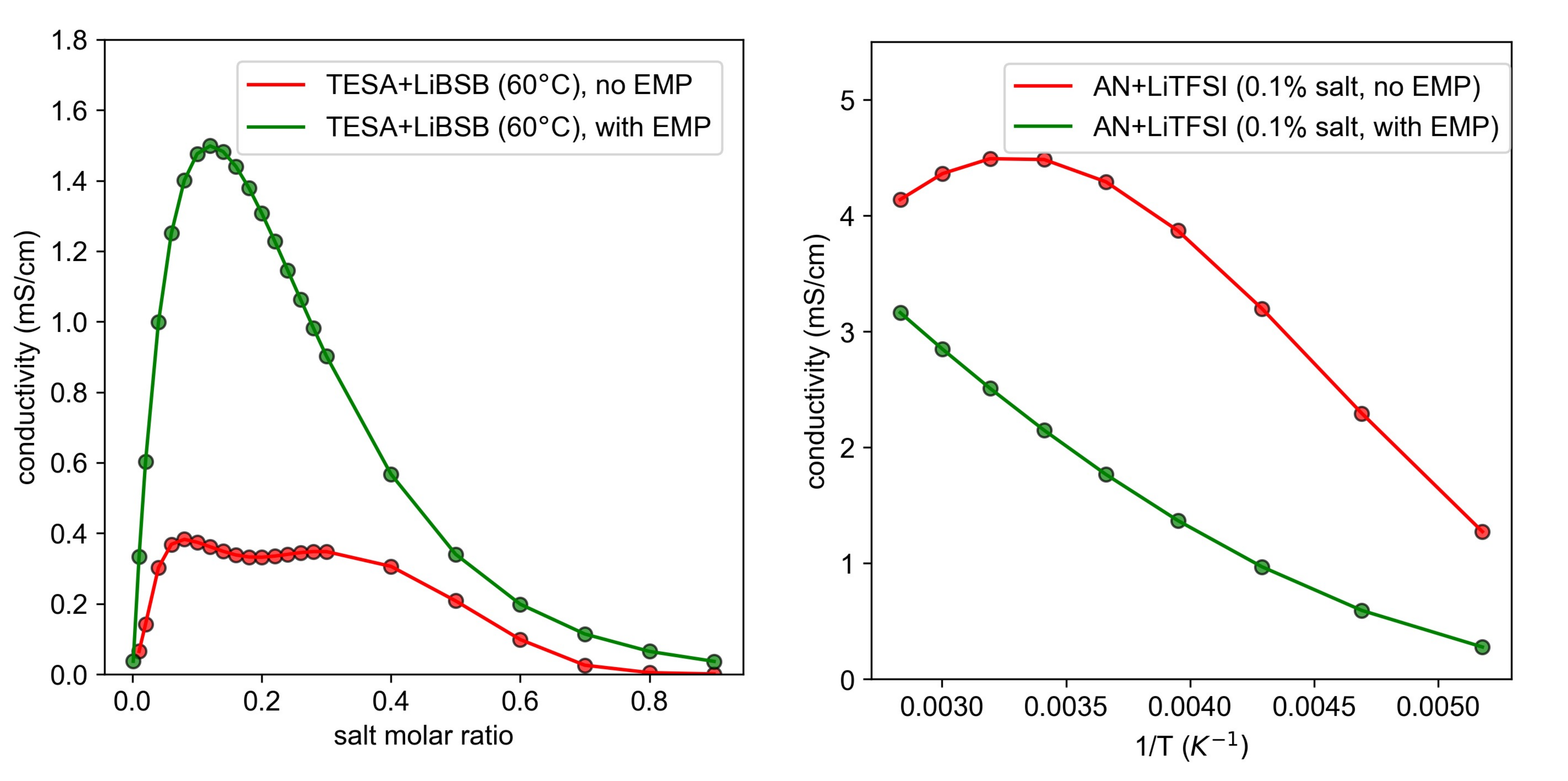}
\caption{\textbf{Unphysical predictions without empirical relation.} The figures show example outliers predicted by model without empirical relation in comparison with our model with empirical equations. \enquote{EMP} stands for \enquote{empirical relation}. The left figure demonstrates a outlier case where conductivities show multiple peaks as concentration varies, while the right figure depicts an incorrect temperature dependence of conductivity as conductivity should increase monotonically with temperature. } 
\label{si_fig:outlier}
\end{figure}

\begin{figure}[htbp!]
\centering
\includegraphics[width=0.85\linewidth]{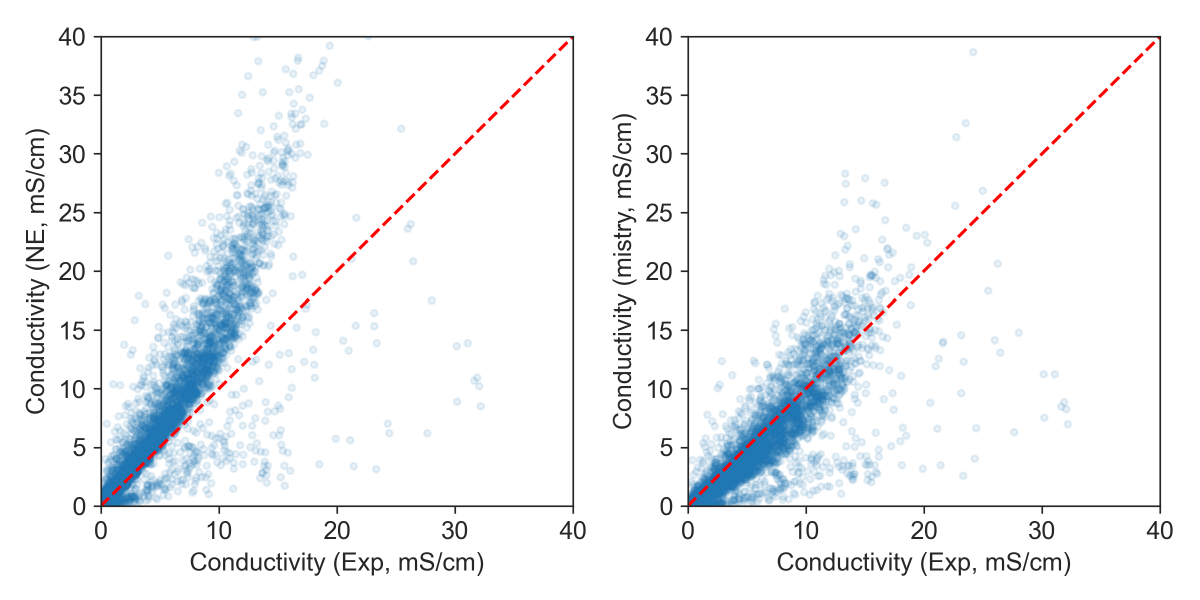}
\caption{\textbf{Comparison of MD conductivities with experimental measurements.} The left parity plot compares NE conductivity with experimental ground truth and the right parity plot compares mistry conductivity.}\label{si_fig:cond_compare}
\end{figure}

\begin{figure}[htbp!]
\centering
\includegraphics[width=0.99\linewidth]{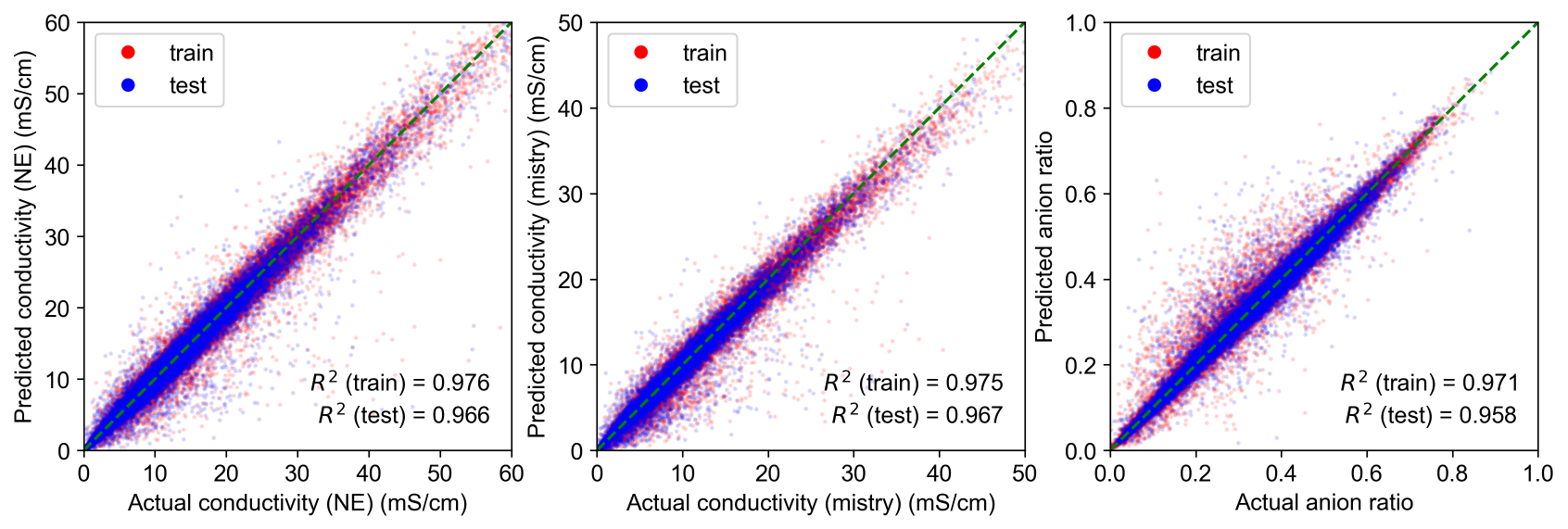}
\caption{\textbf{Predicted electrolyte properties from computational pretraining.} The parity plots comparing conductivities and anion ratio predicted from our model and ground truth from MD simulations. }\label{si_fig:comp_pretrain}
\end{figure}

\begin{figure}[htbp!]
\centering
\includegraphics[width=\linewidth]{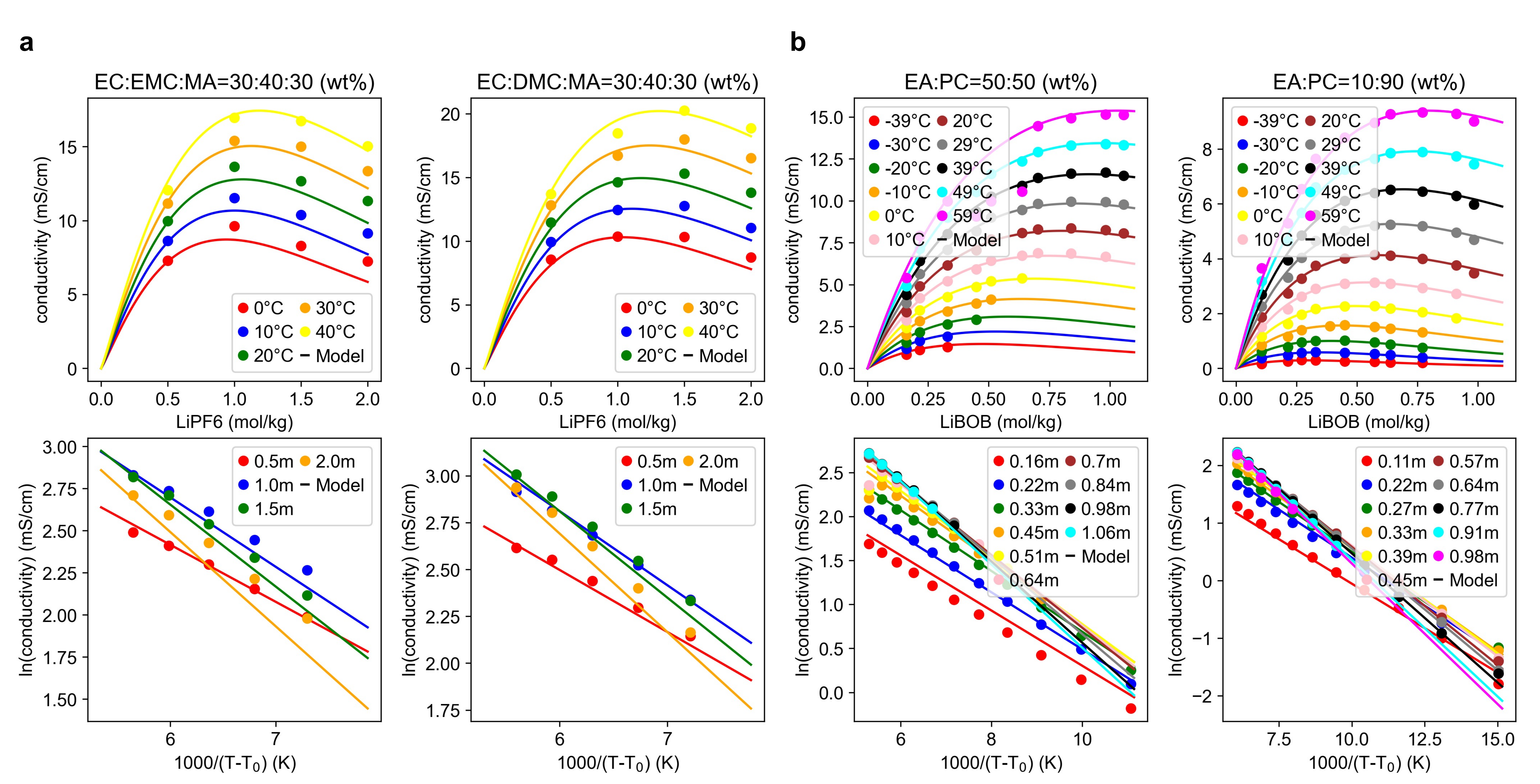}
\caption{\textbf{More results of model predictions on temperature and concentration dependence of conductivity.} \textbf{a} EC/DMC/MA/$\text{LiPF}_6$ and EC/EMC/MA/$\text{LiPF}_6$ systems. The dots are experimental data from Ref.~\protect\citeSI{Logan2018} \textbf{b} EA/PC/LiBOB system. The experimental data are from Ref.~\protect\citeSI{Ding2005}.} 
\label{si_fig:tmdep}
\end{figure}

\begin{figure}[htbp!]
\centering
\includegraphics[width=\linewidth]{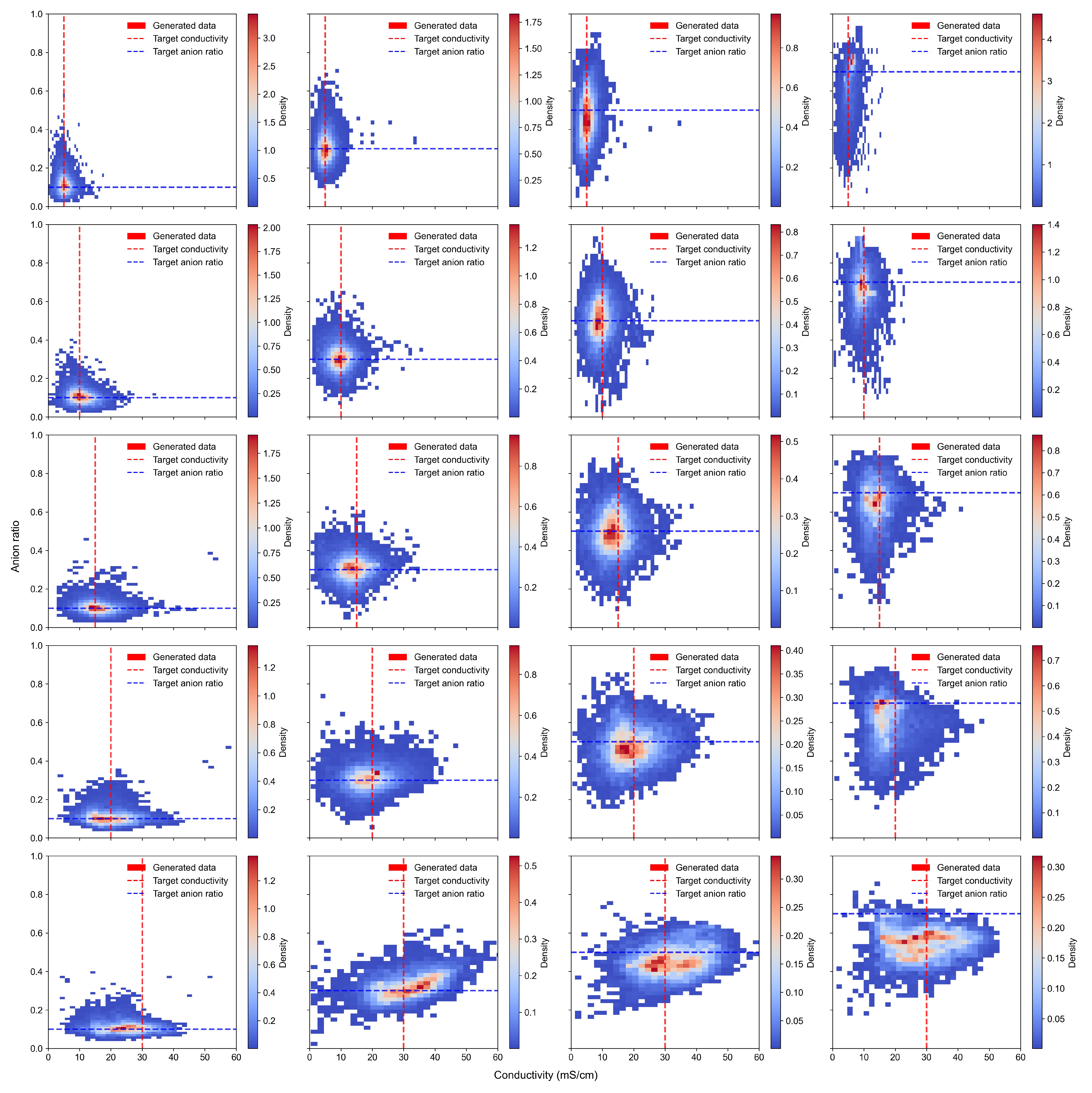}
\caption{\textbf{More results of conditional generation.} Different combinations of target conductivities (5.0, 10.0, 15.0, 20.0, 30.0 mS/cm) and anion ratios (0.1, 0.3, 0.5, 0.7) are tested.}
\label{si_fig:cond_generation}
\end{figure}

\begin{figure}[htbp!]
\centering
\includegraphics[width=\linewidth]{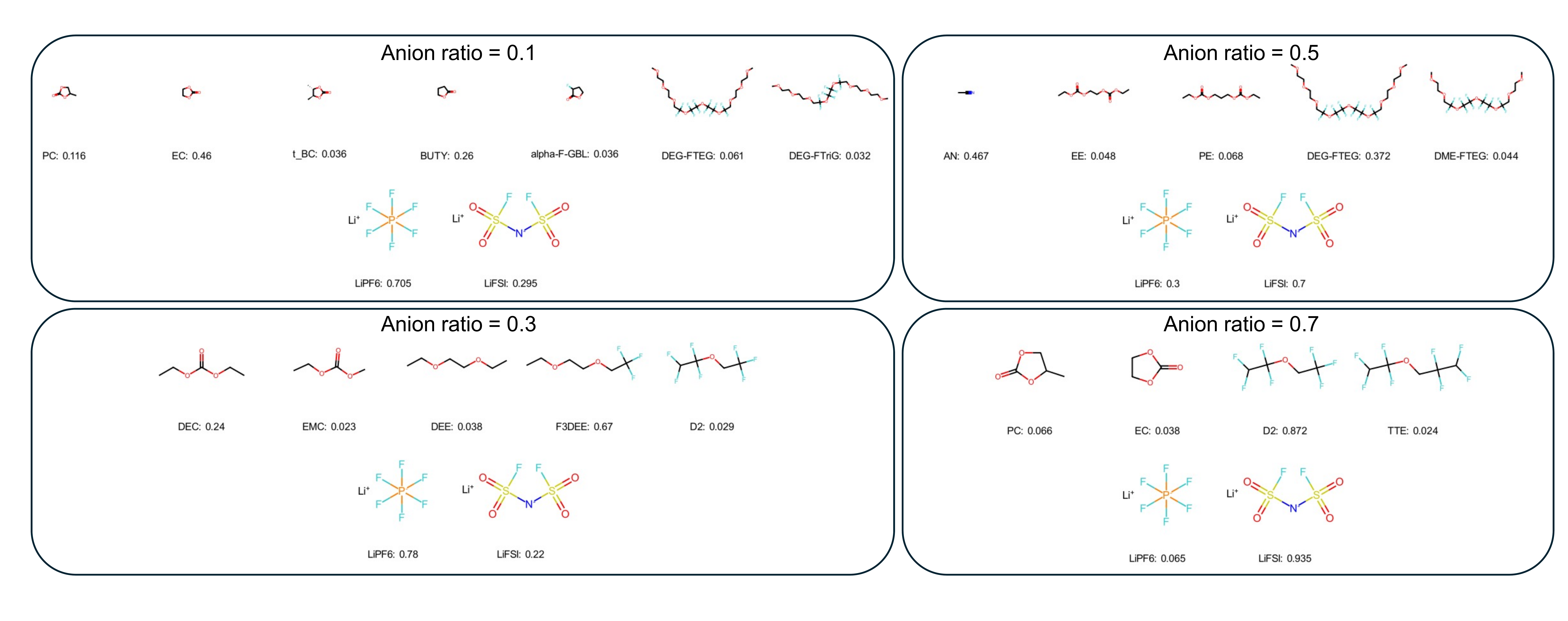}
\caption{\textbf{Example of generated electrolyte formulation (conductivity = 5.0 mS/cm).}}
\label{si_fig:example_formulation_5.0}
\end{figure}

\begin{figure}[htbp!]
\centering
\includegraphics[width=\linewidth]{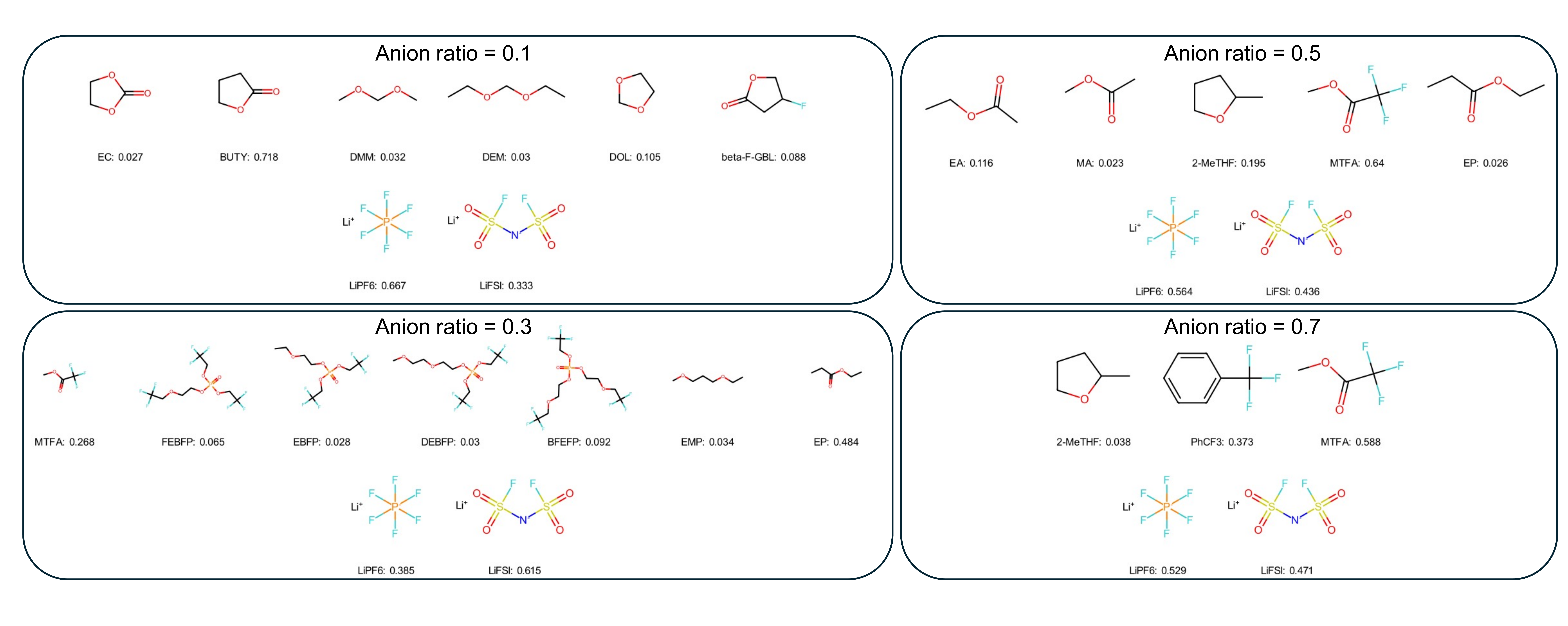}
\caption{\textbf{Examples of generated electrolyte formulation (conductivity = 10.0 mS/cm).}}
\label{si_fig:example_formulation_10.0}
\end{figure}

\begin{figure}[htbp!]
\centering
\includegraphics[width=\linewidth]{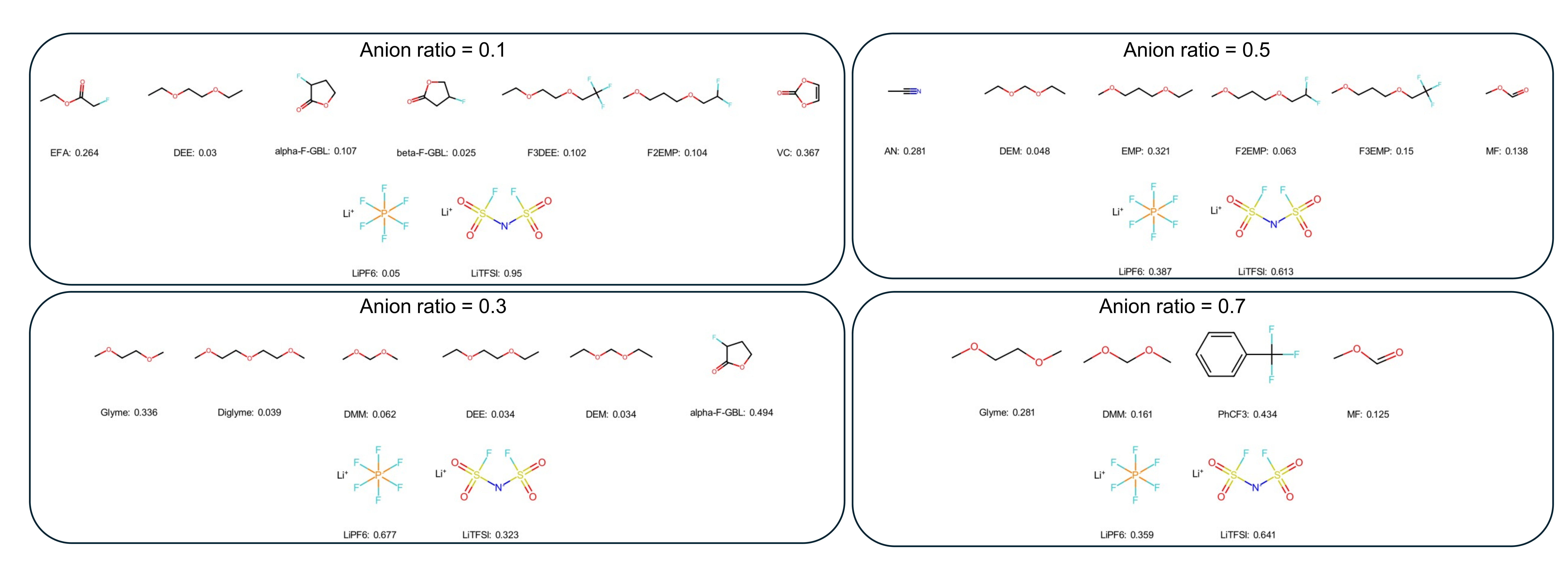}
\caption{\textbf{Examples of generated electrolyte formulation (conductivity = 15.0 mS/cm).}}
\label{si_fig:example_formulation_15.0}
\end{figure}

\begin{figure}[htbp!]
\centering
\includegraphics[width=\linewidth]{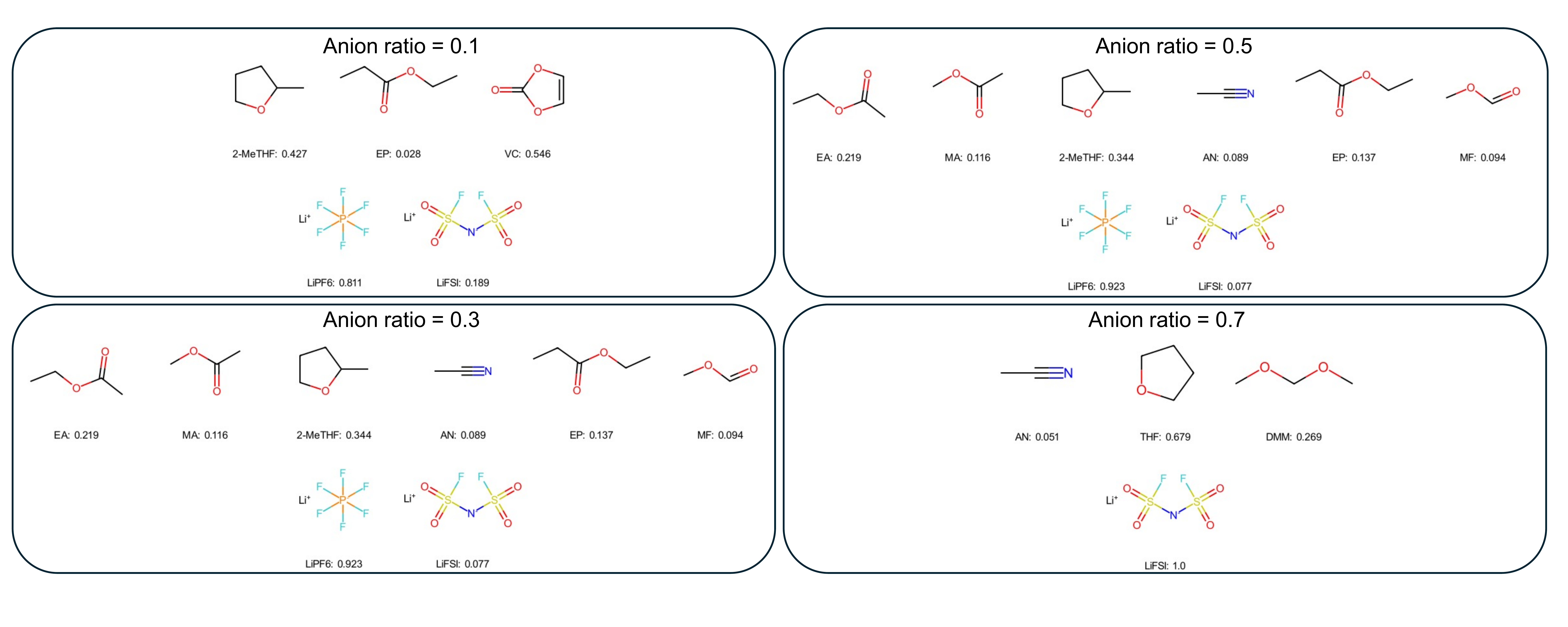}
\caption{\textbf{Examples of generated electrolyte formulation (conductivity = 20.0 mS/cm).}}
\label{si_fig:example_formulation_20.0}
\end{figure}

\begin{figure}[htbp!]
\centering
\includegraphics[width=\linewidth]{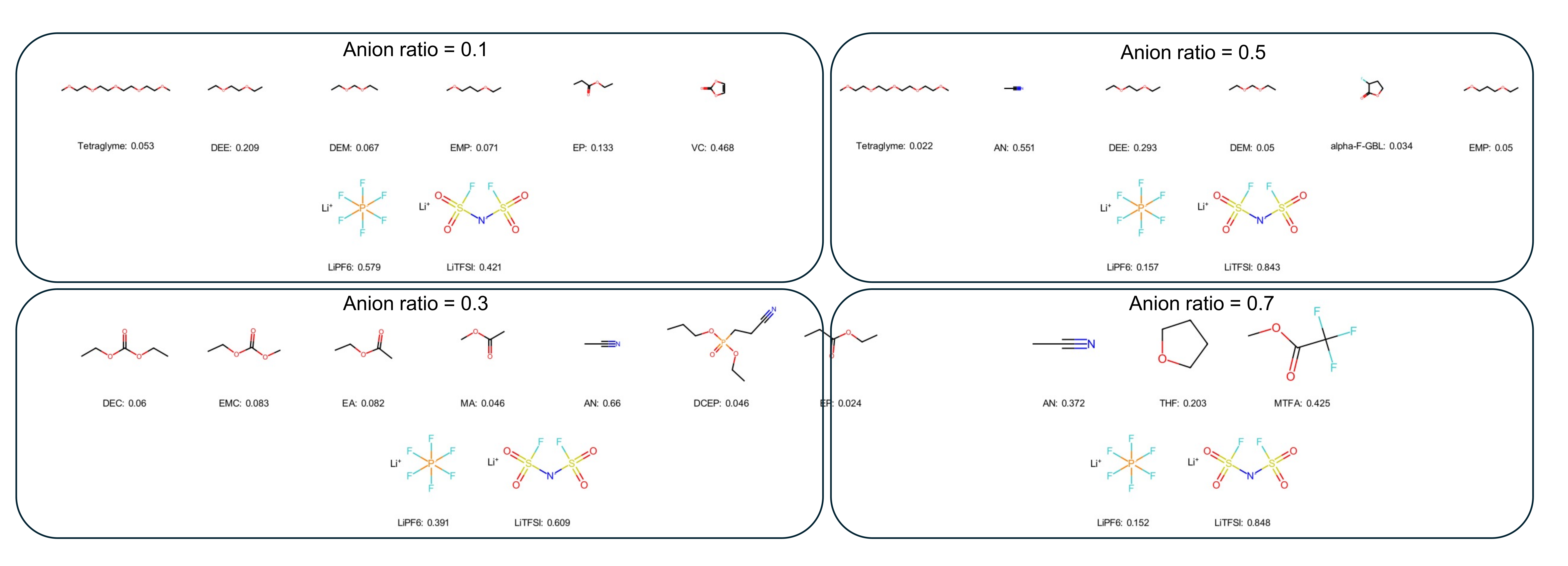}
\caption{\textbf{Examples of generated electrolyte formulation (conductivity = 30.0 mS/cm).}}
\label{si_fig:example_formulation_30.0}
\end{figure}

\clearpage

\end{document}